# Determination of dynamic flow stress equation based on discrete experimental data: Part 1 Methodology and the dependence of dynamic flow stress on strain-rate

Xianglin Huang, Q.M.Li*

School of Engineering, The University of Manchester, Manchester, M13 9PL, UK

* Corresponding author: e-mail, Qingming.li@manchester.ac.uk

Abstract

In this study, a framework to determine the dynamic flow stress equation of materials based on discrete data of varied (or instantaneous) strain-rate from split Hopkinson pressure bar (SHPB) experiments is proposed. The conventional constant strain-rate requirement in SHPB test is purposely relaxed to generate rich dynamic flow stress data which are widely and diversely distributed in plastic strain and strain-rate space.

Two groups of independent SHPB tests, i.e. Group A (without shaper) and Group B (with shaper) were conducted on the C54400 phosphor-bronze copper alloy at room temperature, obtaining flow stress data (FSD) (two-dimensional (2D) matrix). Data qualification criteria were proposed to screen the FSD, with which qualified FSD were obtained.

The qualified FSD of Group A were coarsely filled with missing data and were reconstructed by the Artificial Neural Network (ANN). As a result, finely-filled FSD of Group A were obtained, which were carefully evaluated by the qualified FSD of Group B. The evaluation proves the effectiveness of ANN in FSD prediction.

Next, the finely-filled FSD from Group A were decomposed by Singular Value Decomposition (SVD) method. Discrete and analytical flow stress equation ($f(\varepsilon, \dot{\varepsilon})_{ana}$) were obtained from the SVD results. Finally, flow stress equation ($f(\varepsilon, \dot{\varepsilon})_{MJC}$) based on conventional method were established. Five uncertainties inherent in the conventional method in the determination of the flow stress equation were identified. The comparison between $f(\varepsilon, \dot{\varepsilon})_{ana}$ and $f(\varepsilon, \dot{\varepsilon})_{MJC}$ demonstrated the effectiveness and reliability of the flow stress equation obtained from the proposed method.

# 1    Introduction

Determining the dynamic flow stress equation is a key procedure in the determination of the dynamic constitutive equation of a metallic material. A dynamic flow stress equation can be generally expressed by

$$\sigma = f(\varepsilon, \dot{\varepsilon}, T) \qquad (1)$$

as a function of strain ($\varepsilon$), strain-rate ($\dot{\varepsilon}$) and temperature (T), which is usually determined empirically based on discrete data from various material tests. Split Hopkinson pressure bar (SHPB) technique is the most frequently used method to determine strain-rate effect on flow stress [1].

In addition to the stress equilibrium condition for a valid SHPB test, constant (or more accurately, nearly-constant) strain-rate is preferred in a SHPB test. Great efforts have been paid to design the shape of incident pulse in order to achieve constant strain-rate in a SHPB test (see pp. 49-62, Sections 2.4 and 2.5 in [2]). The measured stress-strain curves (data) under constant strain-rate can clearly show the influence of strain-rate on the flow stress. Similarly, stress-strain curves under various constant temperatures are necessary for the construction of dynamic flow stress equation when the thermal effect is important. Three types of material tests, i.e. (i) quasi-static test at strain-rate between $10^{-4} – 10^{-3}$ s$^{-1}$ and room temperature, (ii) quasi-static tests at various constant environmental temperatures, and (iii) dynamic tests at various constant strain-rates and room temperature, have been considered as 'quasi-standard' material tests for the determination of the dynamic flow stress for a given material (see [3-5]).

The preferred constant strain-rate and temperature requirements of material tests are largely related to the Johnson-Cook (J-C) dynamic flow stress equation [6], which decouples effects of strain, strain-rate and temperature using multiplicative functions that can be determined independently by the above-mentioned three types of material tests. This is inevitable since J-C dynamic flow stress equation has been widely adopted in the studies of high strain-rate responses of structures (e.g. it has been cited for more than 30,000 times in past four decades since its publication, which is searched in [7]).

However, it has been shown that J-C type dynamic flow stress equation is only valid when the discrete flow stress data (FSD) matrix can be decompose into a single term by singular value decomposition (SVD) ([7]), which implies that J-C type dynamic flow stress equation is not generally applicable, and therefore, raises a basic question about the legitimacy of the J-C type dynamic flow stress equation and the necessity of the constant strain-rate and temperature requirements. Meanwhile, the current practices to process dynamic flow stress data have following three problems.

**Problem-1: The use of averaged strain-rate**

In the calibration of material parameters in commonly-used dynamic flow stress equations, averaged strain-rate is usually used. However, the experimentally-obtained strain-rate data in a dynamic test are generally non-constant. Fig. 1 shows typical strain-rate data of four materials obtained from SHPB tests without using pulse shaper. It shows that the strain-rate varies with strain significantly in a SHPB test. Although pulse shaper can effectively reduce the strain-rate variation (see Fig. 2 for the strain-rate variations of the same four materials with the use of pulse shaper), the difference between the averaged strain-rate and the maximum/minimum strain-rates are still considerable.

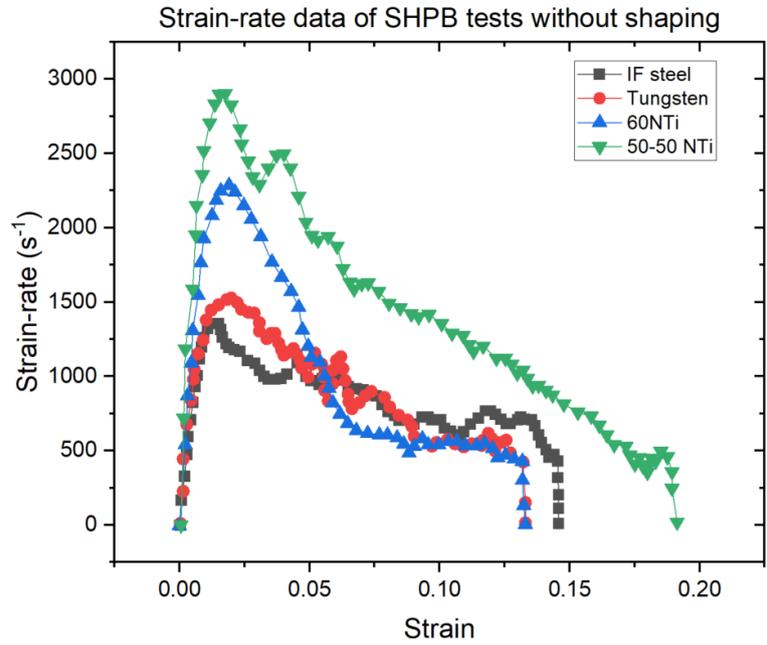

Fig. 1 Typically-recorded strain-rate data of SHPB test reproduced from [8].

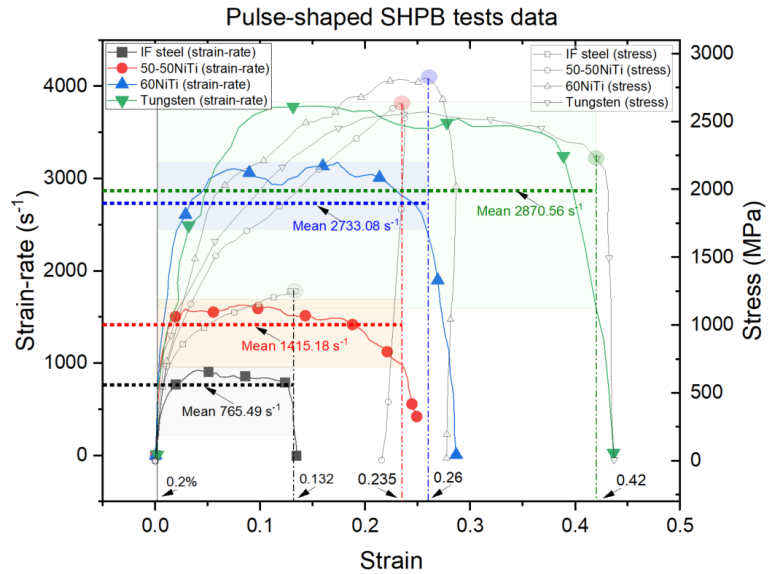

Fig. 2 Strain-rate and stress data (against strain) from SHPB tests with pulse shaper. For each material, the initial yield point is uniformly set at 0.2% plastic strain and the end of strain-rate is defined at the point of the failure strength of each material. Data sources are reproduced from [8].

**Problem-2: Data qualification**

Although the trend of strain hardening effect can be observed, increased stress fluctuation at higher strain-rate is frequently seen (e.g. Fig. 3) where stress at higher strain-rate may be smaller than the stress at lower strain-rate for the same strain, which does not represent the real strain-rate effect on the flow stress. This is caused by the violation of SHPB assumptions (e.g. stress equilibrium, non-unloading), and therefore, the associated data should be disqualified for the material parameter calibration.

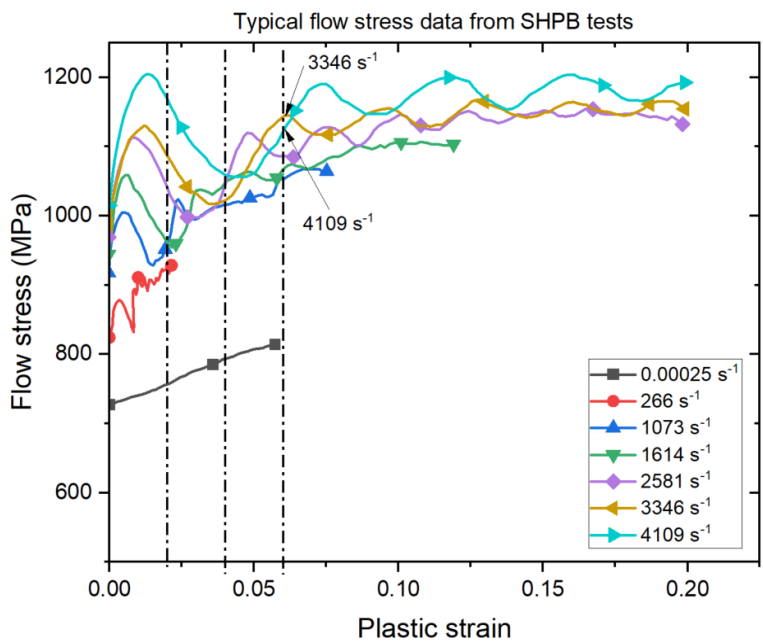

Fig. 3 Dynamic flow stress data reproduced from [9].

**Problem-3: Necessity of using an empirical equation to fit dynamic flow stress data**

Currently-used dynamic flow stress equations are dominantly empirical equations. The accuracy of an empirical dynamic flow stress equation is determined by two factors, i.e. (i) the accuracy of the material testing data, and (ii) the representativeness of the selected empirical function. Factor-(i) is related to the material testing technique, which has been studied extensively (e.g. [1, 2] for SHPB). Logically, if the form of an empirical flow stress equation has none-physical considerations, the choice of a particular flow stress function cannot increase the overall accuracy, i.e. extra errors may be introduced in data-fitting process, and therefore, the overall accuracy is generally less than or equal to the accuracy of the material testing data. The only advantage of using an empirical flow stress function is its analytical expression that can be conveniently described and implemented into numerical codes using conventional method. Such advantage will disappear if material testing data can be directly used through machine learning (ML), which has recently attracted more attentions.

Artificial neural network (ANN), as one of the ML methods, has been widely used in the flow stress determination, e.g. [10-13]. The feature of ANN-based flow stress is that a well-trained ANN network contains significant number of weights and biases. When an ANN-based flow stress is integrated into a numerical analysis program, it requires considerable computational time in each time increment. In addition, the implementation of ANN-based flow stress equation is difficult for users in practical numerical simulations.

In this study, we propose a data-driven method based on the discrete flow stress data (FSD) matrix introduced in [7] to determine the dynamic flow stress equation (Note: this equation can be understood as a complete data set or its interpolated expressions). In the framework of the proposed method, the requirements of constant strain-rate and constant temperature will not be necessary. Instead, varying strain-rate and temperature are preferred in order to obtain more widely-reached (or widely-experienced) material testing data in strain, strain-rate and temperature space. The use of empirical flow stress equation is abandoned to minimising the data-fitting error. Meanwhile, the issue of the material data qualification in SHPB tests will be discussed.

Without losing generality, we focus on the material testing data in two-variable space (i.e. strain and strain-rate space) in this paper so that the framework of the methodology can be introduced and demonstrated with minimum complexity. The same concept and the similar method can be extended to 3-variable space when temperature variation is included, which will be carefully developed in Part 2 of the study in the companion paper [14]. In Section 2, the framework of the methodology is proposed where original non-constant strain-rate data are firstly screened by a qualification criterion so that only qualified material testing data are kept. Then, instead of averaging the varied strain-rates in each test, the qualified material testing data from each test are all directly involved in the determination of the dynamic flow stress equation. In this framework, the qualified material testing data are used to train an ANN to predict dynamic flow stress in order to form a FSD matrix (two-dimensional, or 2D) in the strain and strain-rate space. Finally, SVD is used to decompose the FSD matrix and obtain the discrete dynamic flow stress equation (one-dimensional, or 1D) of the material. With a proper fitting method, analytical constitutive equation can also be obtained.

The above framework is verified and validated by diversified experimental data of SHPB tests. This method can also be easily extended to other dynamic material tests for the determination of the dynamic flow stress of metallic materials.

## 2   Methodology

In this section, a data-driven approach is proposed to obtain the dynamic flow stress equation based on dynamic material testing data. Only strain and strain-rate variables are considered here. As shown in Fig. 4(a), the dynamic flow stress data can be obtained from dynamic material tests for various strains and strain-rates, which can be expressed in a FSD matrix [7]. If these data are generated by SHPB test, they need to be qualified by assessing the stress equilibrium and non-unloading requirements to obtain the qualified FSD matrix (Fig.4(b)). For other dynamic material tests, it is necessary to ensure that the measured data are qualified as material data, which will not be discussed here because it depends on the specific dynamic material testing technique. Qualified FSD matrix only provides a coarsely-filled variable space, depending on the actual design of the dynamic material test. In order to obtain a FSD matrix in a finely-filled variable space, an ANN model is trained firstly by the qualified FSD matrix (Fig.4(c)). Then, the trained ANN can be used to predict the FSD matrix in the full variable space (Fig.4(d)). Finally, the procedures developed in [7] can be used to further obtain the dynamic flow stress equation which can be implemented into numerical models (Figs.4(e-i)).

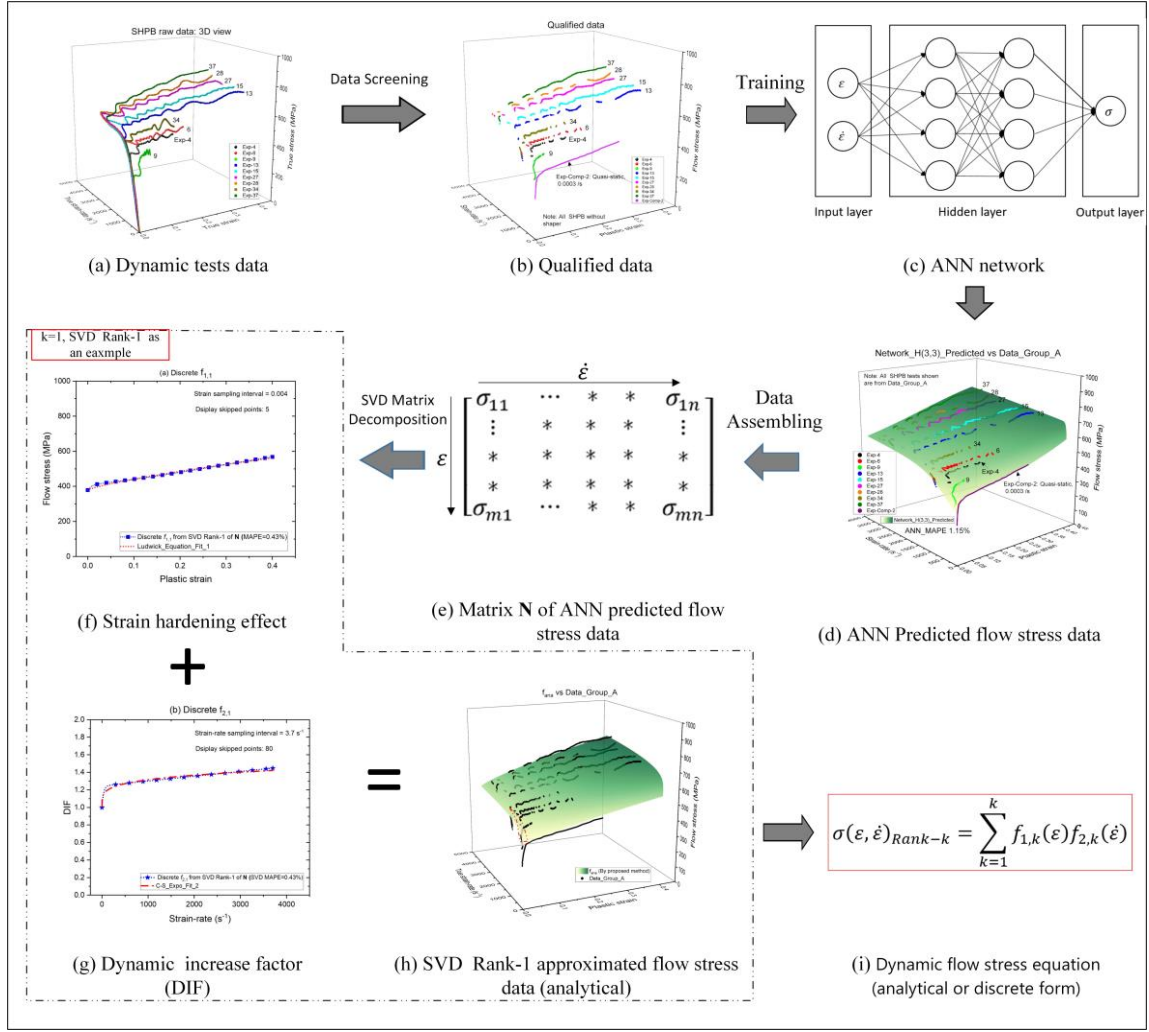

Fig. 4 Schematics of the framework.

## 2.1 Data generation

### 2.1.1 SHPB governing equation

SHPB apparatus is widely used to quantify the dynamic behaviour of materials. When stress equilibrium is met, the engineering stress $\sigma$, strain $e$ and strain-rate $\dot{e}$ in a SHPB specimen can be obtained by

$$\dot{e} = -2\frac{c_b}{L_s}(\varepsilon_r) \qquad (2)$$

$$e = -2\frac{c_b}{L_s}\int_0^t (\varepsilon_r)\,dt \qquad (3)$$

$$\sigma_{front} = \frac{A_b}{A_s}E_b(\varepsilon_i + \varepsilon_r) \qquad (4)$$

$$\sigma_{rear} = \frac{A_b}{A_s} E_b \varepsilon_t \tag{5}$$

where $\varepsilon_i$ and $\varepsilon_r$ are the incident and reflected strain signals recorded from the incident bar, respectively; $\varepsilon_t$ is the transmitted strain signal recorded from the transmitter bar; $c_b$ is the 1D sound speed in the pressure bar; $L_s$ is the length of the specimen; $A_b$ and $A_s$ are the cross-sectional areas of the pressure bar and the specimen, respectively; $E_b$ is the elastic module of the pressure bar; $\sigma_{front}$ and $\sigma_{rear}$ are the engineering front stress (stress between the interface of incident bar and specimen) and engineering rear stress (stress between the interface of specimen and transmitter bar), respectively.

True strain, true strain-rate and true stress can be related to engineering strain, engineering strain-rate and engineering stress in uniaxial compression state under the assumption of plastic incompressibility, i.e.

$$\varepsilon = -\ln(1 - e) \tag{6}$$

$$\dot{\varepsilon} = \frac{d\varepsilon}{dt} = \frac{\dot{e}}{1 - e} \tag{7}$$

$$\sigma = \sigma_e (1 - e) \tag{8}$$

where $\varepsilon$, $\dot{\varepsilon}$ and $\sigma$ are true strain, true strain-rate and true stress, respectively. Finally, the plastic strain, strain-rate and flow stress can be obtained by deducting their elastic counterparts.

2.1.2 Data qualification

Although, true strain, true strain-rate and true stress data are obtained from above equations, some data are not qualified for the determination of the dynamic flow stress. Therefore, further assessment on these data should be made to obtain qualified dynamic flow stress data. For SHPB tests of metals, if the pressure bar diameter is small and the optimal length/radius ratio is used for the specimen, the main qualification check is the stress equilibrium in the specimen.

Stress equilibrium (uniformity) in a SHPB specimen, which has been widely investigated (e.g. [15, 16]), is used in this study as a criterion to identify the non-equilibrium data, i.e.

$$e_{equi} = \left|\frac{\sigma_{\text{front}} - \sigma_{\text{rear}}}{\sigma_{\text{fornt}}}\right| < e_{cr} \tag{9}$$

where $e_{equi}$ is the stress equilibrium measure and $e_{cr}$ is a selected threshold value. When $e_{equi} < e_{cr}$, the qualified stress data can be obtained.

In addition, the instantaneous strain-rate should be positive to ensure that the specimen is not undergoing unloading, i.e.

$$\dot{\varepsilon} > 0 \tag{10}$$

## 2.2 Generation of finely-filled FSD matrix by ANN

When the raw data from SHPB tests are screened by the above criterion, the qualified data of flow stress are coarsely filled in strain and strain-rate space. The coarsely-filled data are then used to train ANN to get finely-filled FSD matrix.

The application of ANN method to the development of material's constitutive equation has been investigated and proved to be effective by many researchers (e.g. [12, 13]). With ANN, the non-linear relationship between stress (output) and other inputs (i.e. strain and strain-rate here) can be established.

The basic ANN procedure used here is briefly described as follows. Before the start of training, the measured input data (strain, strain-rate) and measured output data (flow stress) should be normalized and projected into the range of [-1, +1]. For a variable $x_i$ ($\varepsilon$ or $\dot{\varepsilon}$) of the input in Fig.5, the normalized $x_i$ is given by

$$\bar{x}_i = 2\frac{x_i - x_i^{\min}}{x_i^{\max} - x_i^{\min}} - 1, i = 1,2 \tag{11}$$

where $x_i^{\min}$ and $x_i^{\max}$ are the minimum and maximum values of $x_i$. The output data y can be normalised in the same way by replace $x_i$ and $\bar{x}_i$ to y and $\bar{y}$, and replace $x_i^{\min}$ and $x_i^{\max}$ by $y_i^{\min}$ and $y_i^{\max}$.

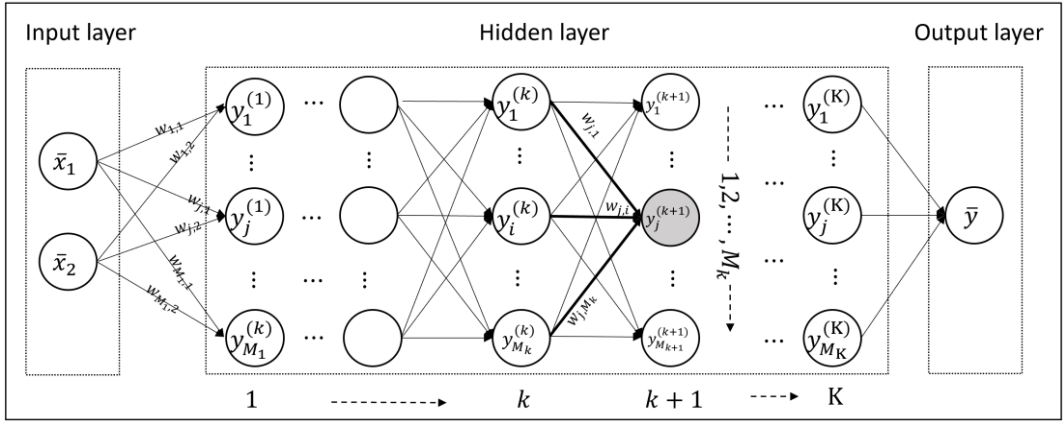

Fig. 5 ANN structure of K hidden layers.

Then, the neuron value of the first hidden layer (k=1) is calculated by

$$y_j^{(1)} = F^{(1)}\left(\sum_{i=1}^{M_0} w_{ji}^{(1)} \bar{x}_i + b_j^{(1)}\right), \qquad (j = 1, \cdots, M_1) \qquad (12)$$

where $M_0 = 2$ is the number of neurons in the input layer; $w_{ji}^{(1)}$ are the weights connecting the neurons in the input layer and the first hidden layer; $b_j^{(1)}$ is the bias of the $j$th neuron in the first hidden layer; $F^{(1)}(\ )$ is the activation function (e.g. Sigmoid, Gaussian) which controls the output of each neuron in the first hidden layer.

The computation of the $j$th neuron in the subsequent $k$th hidden layers is conducted in a similar way as follows

$$y_j^{(k)} = F^{(k)}\left(\sum_{i=1}^{M_{k-1}} w_{ji}^{(k)} y_i^{(k-1)} + b_j^{(k)}\right), \qquad (j = 1, \cdots, M_k; 2 \leq k \leq K) \qquad (13)$$

where $M_k$ represents the number of neurons in the $k$th hidden layer.

Finally, the value of the output neuron $\bar{y}$ is computed from the last hidden layer.

$$\bar{y} = F^{(K+1)}\left(\sum_{i=1}^{M_K} w_{1i}^{(K+1)} y_i^{(K)} + b_1^{(K+1)}\right) \qquad (14)$$

where $w_{1i}^{(K+1)}$ are the weights connecting the neurons in the $K$th layer and the output (i.e. (K+1)th) layer; $b_1^{(K+1)}$ is the bias of the output neuron; $M_K$ is the number of neurons in the last ($K$th) hidden layer. Conducting the inverse normalisation of output $\bar{y}$, the predicted flow stress $\sigma$ can be obtained by

$$\sigma = y = \frac{(\bar{y}+1)y^{max} - (\bar{y}-1)y^{min}}{2} \tag{15}$$

where $y^{min}$ and $y^{max}$ are the minimum and maximum values of the predicted flow stress data (output).

2.3 Finely-filled FSD matrix decomposition and flow stress determination

Now, for a preferred mesh grid of strain and strain-rate, the whole field discrete flow stress vs (strain, strain-rate) data are obtained. The 2D FSD matrix $\mathbf{N}$, with the element of $\mathbf{N}$ denoted as $N_{i,j} = \sigma_{i_1 i_2}$ ($0 < i_1 \leq m = I_1, 0 < i_2 \leq n = I_2$), can be assembled as

$$\mathbf{N} = \begin{array}{c} \\ \varepsilon_1 \\ \downarrow \\ \varepsilon_m \end{array} \begin{array}{c} \dot{\varepsilon}_1 \rightarrow \dot{\varepsilon}_n \\ \begin{bmatrix} \sigma_{11} & \cdots & \sigma_{1n} \\ \vdots & \ddots & \vdots \\ \sigma_{m1} & \cdots & \sigma_{mn} \end{bmatrix} \end{array} \tag{16}$$

To reduce the dimension of $\mathbf{N}$ or to obtain an analytical equation of $\sigma(\varepsilon, \dot{\varepsilon})$, $\mathbf{N}$ can be further decomposed by SVD method as reported in [7]. The method is briefly described below.

According to Eckart-Young theorem [17], a matrix $\mathbf{N}$ can be decomposed into R terms and approximated as follows

$$\begin{aligned} \mathbf{N}^{I_1 \times I_2} &= \sum_{k=1}^{R} \mathbf{N}_k^{I_1 \times I_2} = \mathbf{U \Lambda V}^T \\ &= \sum_{k=1}^{R} \lambda_k \mathbf{U}(:,k) \otimes \mathbf{V}(:,k) \\ &\approx \sum_{k=1}^{r} \lambda_k \mathbf{u}_k \otimes \mathbf{v}_k \end{aligned} \tag{17}$$

where $\mathbf{U} = (\mathbf{u}_1, \mathbf{u}_2, \cdots, \mathbf{u}_R)$, $\mathbf{\Lambda} = \begin{bmatrix} \lambda_0 & \cdots & \mathbf{0} \\ \vdots & \ddots & \vdots \\ \mathbf{0} & \cdots & \lambda_k \end{bmatrix}$ and $\mathbf{V} = (\mathbf{v}_1, \mathbf{v}_2, \cdots, \mathbf{v}_R)$ are $I_1 \times I_1$ orthogonal matrix, $I_1 \times I_2$ diagonal matrix and $I_2 \times I_2$ orthogonal matrix, respectively. $\lambda_k$ is the $k$th non-zero singular value of $\mathbf{\Lambda}$. In each term in Eq.(17), the strain and strain-rate effects are decoupled. Let $f_{1,k}((x_1)_i) = \lambda_k \mathrm{u}_k(i)\mathrm{v}_k(1)$ and $f_{2,k}((x_2)_j) = \frac{\mathrm{v}_k(j)}{\mathrm{v}_k(1)}$, we have

$$f(\varepsilon, \dot{\varepsilon}) \approx \sum_{k=1}^{r} f_{1,k}(\varepsilon) f_{2,k}(\dot{\varepsilon}) \tag{18}$$

where $\mathrm{u}_k(i)$ is the $i$th element of vector $\mathbf{u}_k$ and $\mathrm{v}_k(j)$ is the $j$th element of vector $\mathbf{v}_k$.

An analytical formula of dynamic flow stress equation can be obtained by selecting proper analytical fitting functions of $f_{1,k}$ and $f_{2,k}$ (e.g. using J-C type decoupled functions) for $r$ additive terms. The error mainly comes from the fitting functions of $f_{1,k}$ and $f_{2,k}$.

In addition, interpolation method can be used to avoid the procedure of finding proper analytical fitting functions. The $f_{1,k}$ and $f_{2,k}$ in Eq.(18) can be obtained from interpolating $\mathbf{u}_k$ and $\mathbf{v}_k$ by proper interpolation method such as linear and spline interpolations. The error mainly comes from the interpolation, which however can be controlled easily by refining the strain and strain-rate grids.

The mean absolute percentage error (MAPE) is introduced to indicate the residual error introduced from the whole procedure.

$$\mathrm{MAPE} = \frac{100\%}{n} \sum_{1}^{n} \left| \frac{\sigma_e - \sigma_m}{\sigma_e} \right| \tag{19}$$

where $\sigma_e$ and $\sigma_m$ are the actual (experimental) and modelling (predicted) values, respectively; n is the total number of points in $\sigma_e$.

## 3    Experimental set-up and results

The material used in this study is C54400 phosphor bronze-copper alloy, which is widely used in the electrical products. The element compositions provided by the supplier are given in Table A.1 in Appendix A. The C54400 alloy has good wear resistance and high stiffness due to the existence of phosphorus while its corrosion resistance and strength are improved by the existence of element Tin (Sn). Meanwhile, the C54400 alloy has excellent formability and solderability.

## 3.1 Quasi-static compressive test at normal temperature (293 K)

Instron universal machine (Maximum capacity of loading: 100 kN) is used to compress cylindrical specimens (see Fig. 6(a) under quasi-static state (0.0003 s$^{-1}$). The gauge length is 10 mm, while the crosshead loading speed is 0.18 mm/min. A low-speed camera (type: Prosilica GT) is installed in front of the specimen to capture its deformation. Virtual extensometer is used to obtain the strain of the specimen by analysing the images obtained from the camera, as shown in Fig. 6(b).

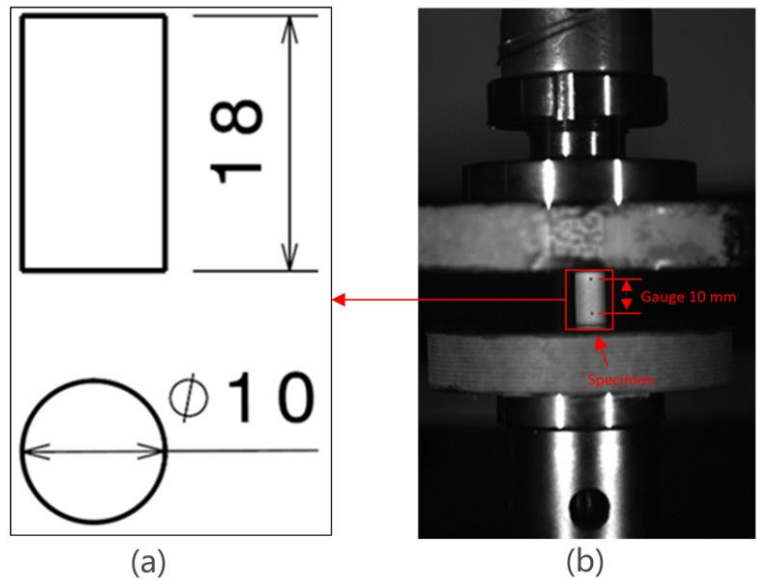

Fig. 6 (a) Geometry of the compression test specimen (units in millimetre), (b) Image taken from the low-speed camera.

The quasi-static testing results of two repeated compression tests are shown in Fig. 7. Young's module E is obtained from the true stress-strain data ($\sigma_{true}, \varepsilon_{true}$) shown in Fig. 7(a). Then the true plastic strain shown in Fig. 7(b) is obtained based on the following equation

$$\varepsilon_p = \varepsilon_{true} - \frac{\sigma_{true}}{E} \qquad (20)$$

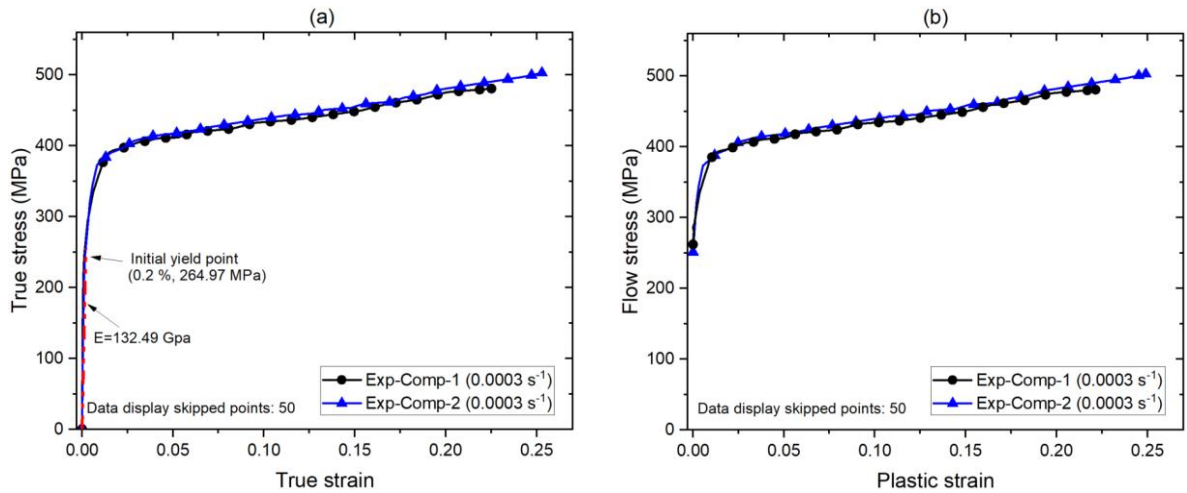

Fig. 7 Quasi-static compressive test results: (a) true total strain, and (b) true plastic strain data. Note: 'Data display skipped points: 50' means that 50 points were skipped between any two consecutive points shown for concise display purpose. This 'skipped' explanation applies to the other figures in this study.

### 3.2 SHPB set-up

To characterize the dynamic behaviour of C54400 copper alloy, a SHPB apparatus was used. The pressure bar diameters are 19.3 mm and made from quenched high strength stainless steel. The Young's modulus of the steel is 206 GPa while the density is 7900 kg/m$^3$. The geometrical dimensions of the specimen are shown in Fig. 8 (Left). The smoothness and parallelism of the specimen surfaces are checked by the excellent fit between two machined specimens, as shown in Fig. 8 (Middle). A well-placed specimen between the two SHPB pressure bars is also shown in Fig. 8 (Right).

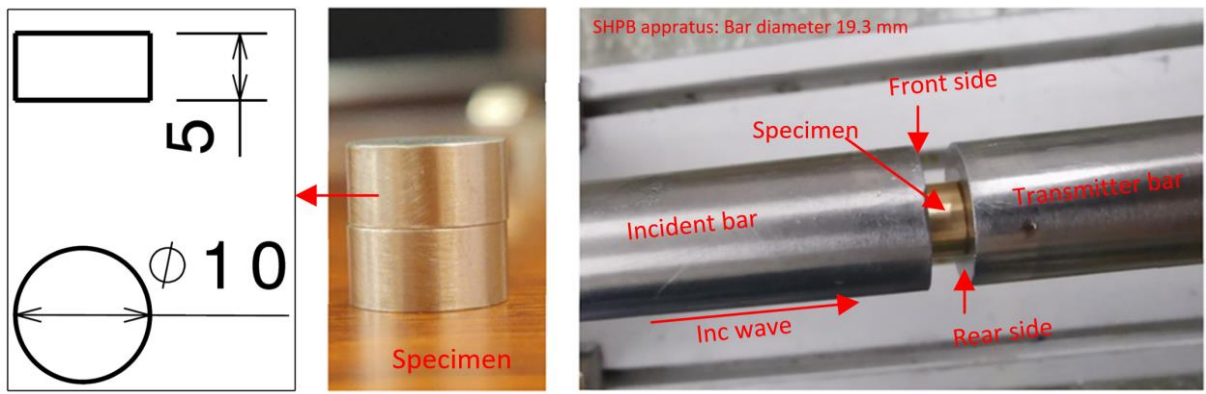

Fig. 8 Geometries of SHPB test specimen (Left), machined specimen (Middle) and the specimen placed between the SHPB pressure bars (Right).

Pulse shapers made from nylon of different thickness and copper alloy were used to achieve diversified signals in SHPB tests, as shown in Fig. 9. The nylon shapers have uniform diameter of 16 mm and the handwritten numbers show the thickness (in mm). The diameter and thickness of the copper alloy are 19 mm and 1 mm, respectively.

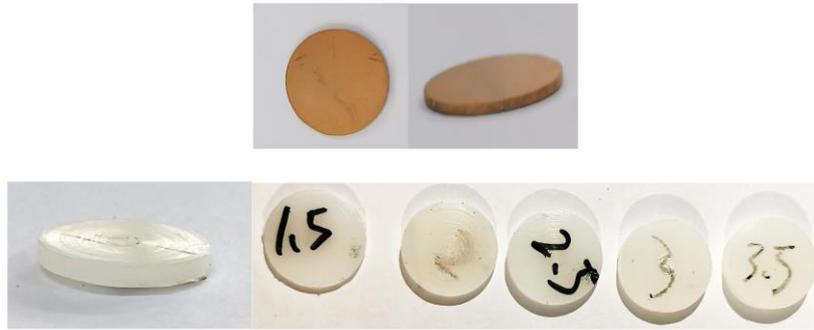

Fig. 9 Copper alloy (upper) and Nylon (lower) shapers used.

### 3.3 SHPB experimental results

In this study, all SHPB tests were conducted at room temperature (293 K). The experimental log is presented in Table A.2, which includes the impact velocities of the striker bar, length of the striker bar, type of shaper and so on. The raw SHPB strain signals recorded by an oscilloscope during the test are shown in Fig. 10. The sampling rate of the oscilloscope is 12.5 MS/s (or 80 ns/pt). Ch1 (Channel 1) and Ch2 (Channel 2) of the oscilloscope are signals from the strain gauges located at incident and transmitter bars, respectively. The geometry of SHPB apparatus is shown in the insert of Fig. 10 (e). The distance of the strain gauge with respect to the specimen/bar interfaces and other SHPB dimensions are presented.

In Fig. 10(a, b), the presented signals are from tests without shaper. Two striker bars with different lengths were used to create different impact durations and velocities. Severe oscillations are observed at the end of the initial rising time of the incident and reflected waves. In contrast, the transmitted waves have less oscillation and are smooth due to the filtering of high frequency wave components by the specimen.

Shapers were used for the tests shown in Fig. 10 (c, d, e). It shows that the wave-shaper affects the profiles of the incident and reflected waves noticeably. The application of Nylon shaper reduces the oscillation of the incident wave at the end of the initial rising stage. However, after the crushing of the Nylon shaper, the 'buffer' between the striker and the incident bar disappears, causing a second loading stage.

Such second loading is conventionally avoided by further increase the thickness of shaper or use other shaper materials, because the difference of strain-rates created at the two loading stages is large, which would cause significant error if averaged strain-rate method was used in the flow stress determination. However, in this study, the second loading is purposely created and the generated flow stress data are kept, which can help to adjust the flow stress data distribution (or data structure) in the strain and strain-rate space, diversify the domain of the dynamic flow stress data, and verify the obtained flow stress equation.

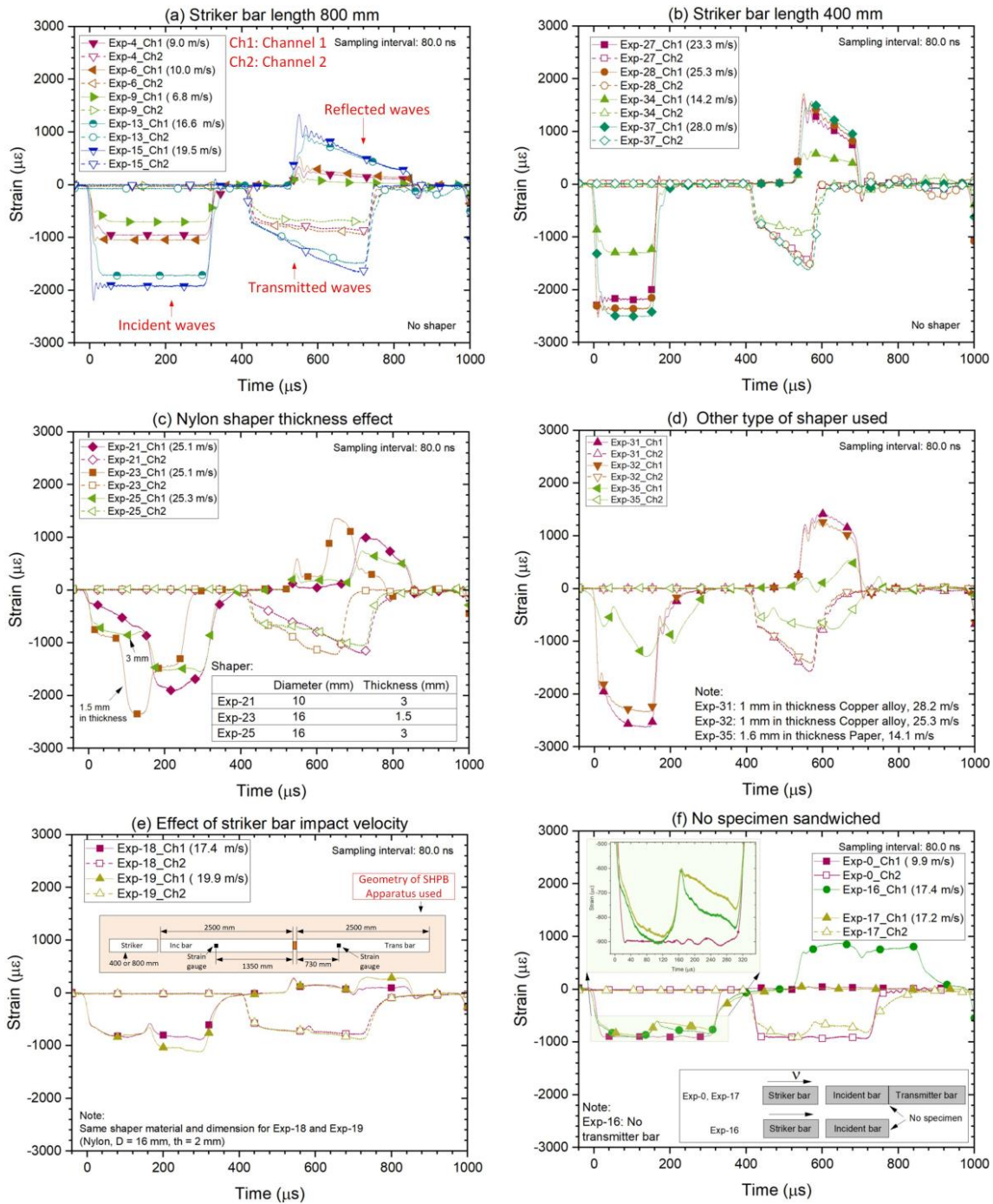

Fig. 10 SHPB raw signals: (a, b) tests without shaper, (c) tests with nylon shapers in different thickness, (d) tests with other shapers, (e) Tests at different impact velocity with nylon shaper, (f)Tests without specimen. Striker bar length used in (c, d, e, f) is 400 mm.

To investigate the effects of shaper's thickness on the second loading peak of the incident wave, two different thicknesses of nylon shapers are impacted by the striker at the same nominal impact velocity of 25 m/s. In addition, a SHPB test (Exp-28) without shaper is presented for comparison. As shown in Fig. 11, the shaper significantly reduces the first peak of the incident wave for both tests. The second peak decreases with the increase of shaper's thickness while the durations of both first and second plateau loading stages increase with the shaper's thickness. Consequently, diversified reflected wave signals, which is directly related to the instantaneous strain-rate, are produced by the use of shaper (see Fig. 11).

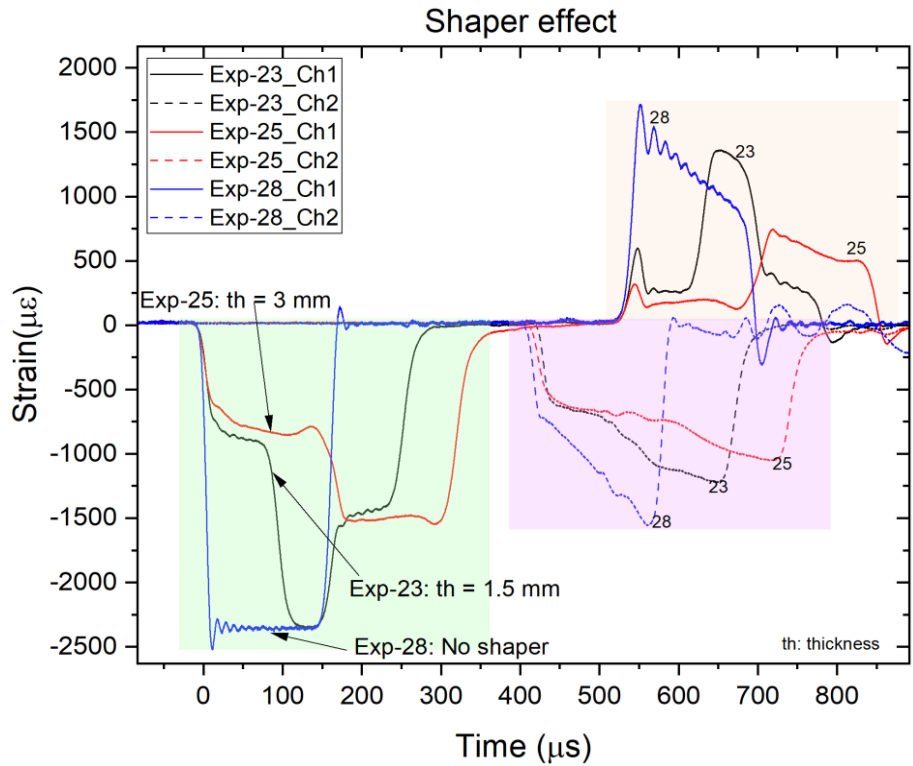

Fig. 11 The effect of nylon shaper thickness on the recorded SHPB waves.

Using the SHPB formulae introduced in Section 2.1.1, rich true stress data in strain and strain-rate space were obtained. The obtained true stress data were divided into two groups, i.e. Group A (without shaper) and Group B (with shaper). Group A is used as ANN training data, while Group B is used as independent testing data for the post-evaluation of the trained ANN.

The common features of Group A data are: 1) relative smooth variation of stress with strain in (stress, strain) space (Fig. 12 (a)); 2) rapid increase and continuous decrease of strain-rate with strain in (strain-rate, strain) space (Fig. 12 (c)); 3) obvious oscillation of stress with strain and strain-rate in (stress, strain, strain-rate) space (Fig. 12 (c)).

In Group B, Exp-32 and Exp-35 (tests with copper alloy shaper) are favourite tests in the conventional determination of the flow stress (e.g. [8, 20]) because their strain-rate curves vary slowly with strain (see Fig. 12 (e)) after the end of initial rising stage and can be treated as constant without causing significant errors. The remaining tests in Group B are conventionally considered as invalid and are avoided experimentally because their strain-rate curves vary significantly with strain (see Fig. 12 (e)). If average strain-rate is used for these tests in Group B, significant information will be lost and considerable errors will incur. In the present study, they are considered as important tests and can be used either in the flow stress determination or in post-evaluation after they pass the data screening criterion proposed in Section 2.1.2. The detailed analysis of Group B data will be presented in the following sections.

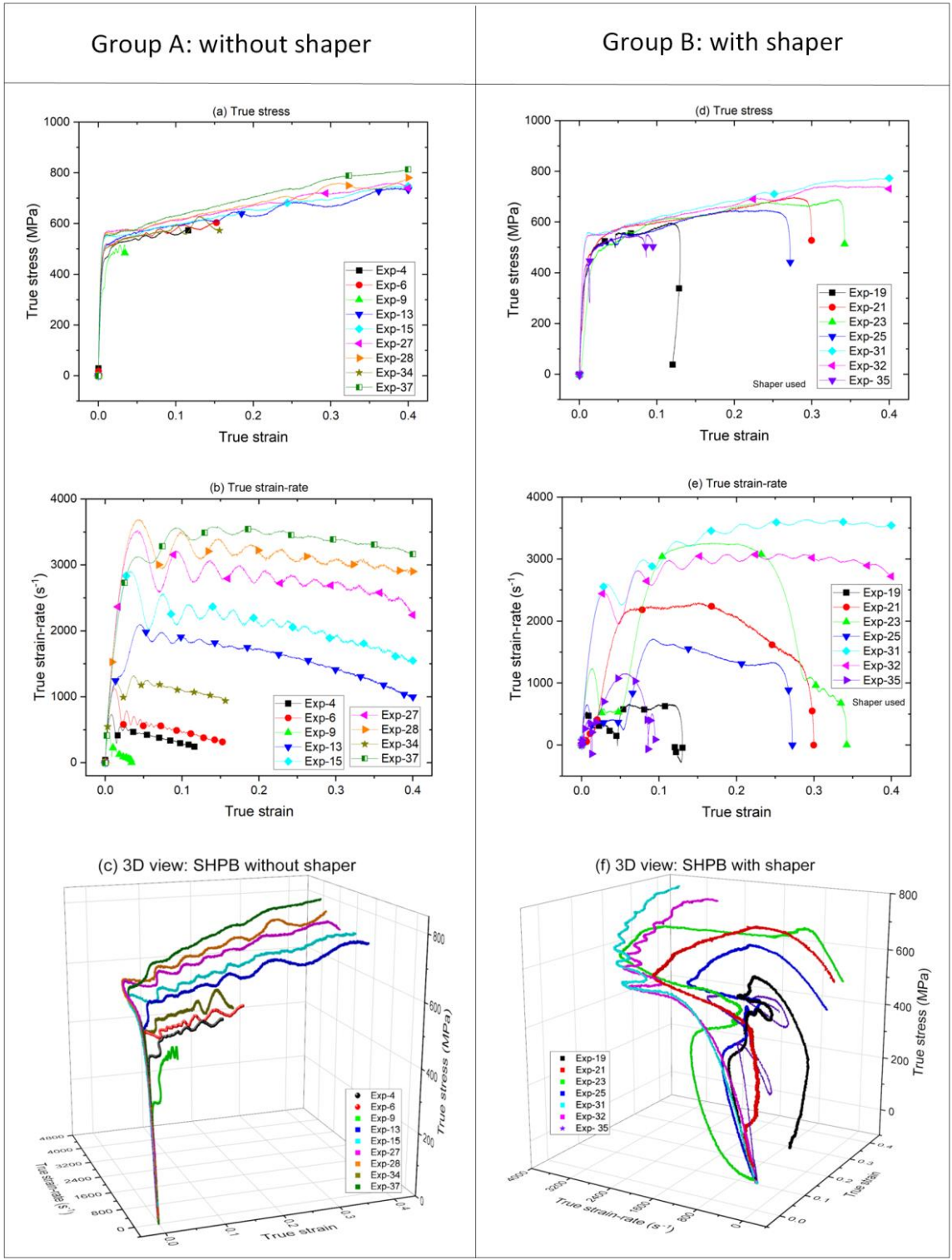

Fig. 12 The obtained true stress data in (stress, strain), (strain-rate, strain) and (stress, strain, strain-rate) spaces: (a, b, c) SHPB tests without shaper (Group A), and (d, e, f) SHPB tests with shaper (Group B). Note: the strain-rate, strain and stress are calculated according to Eqs.(2), (3) and (5), respectively. In addition, no filtering or smoothing technique was used (Note: Exp-18 used later in Section 4.2 is not included in Group B because Exp-19, which has almost the same test condition as Exp-18 is included in Group B).

## 4  Data generation

In this section, the effect of loading rate on stress equilibrium is qualitatively analysed. Evidence of strong instantaneous correlation between flow stress and strain-rate is presented. The data qualification criteria introduced in Section 2.1.2 are implemented to obtain qualified data, which are named as Group_A_Data (from SHPB tests without using shaper, i.e. Group A in Fig. 12) and Group_B_Data (from SHPB tests with using shaper, i.e. Group B in Fig. 12). Group_A_Data are then used in the development of the ANN algorithm (includes training, validating and testing), while the Group_B_Data are used for the post-evaluation of the trained ANN.

### 4.1  Loading rate analysis

The variation of the incident wave ($\varepsilon_{inc}$) with time is an indicator of the loading rate. Stress wave non-equilibrium between the two sides of specimen often occurs when the loading rate is too high.

Fig. 13 shows the stress wave equilibrium errors ($e_{wave}$) and their corresponding signals from four typical SHPB tests. The stress wave equilibrium error is defined as $e_{wave} = \left| \frac{\varepsilon_{inc} - \varepsilon_{ref} - \varepsilon_{trans}}{\varepsilon_{inc} - \varepsilon_{ref}} \right|$, where $\varepsilon_{inc}$, $\varepsilon_{ref}$ and $\varepsilon_{trans}$ are respectively the measured (in absolute value) incident, reflected and transmitter strains in pressure bars.

If the maximum equilibrium error tolerance is set as 5% (see the light-yellow box in Fig. 13), which is sufficient for an engineering analysis, it can be seen that the stress wave in the specimen cannot be balanced at the early loading stage (e.g. time < 90 μs in Fig. 13(a) and time < 100 μs in Fig. 13(b)). When the stress wave equilibrium is achieved, the incident waves shown in Fig. 13(a, b) become flat and the two sides of the specimen are loaded by almost the same stress curves (see the ($\varepsilon_{inc} - \varepsilon_{ref}$) and $\varepsilon_{trans}$ curves in the light-green box in Fig. 13(a,b)). It is noted that the strain-rates continously decrease when the stress wave equilibrium is reached (see the green curves in the light-green box in Fig. 13(a,b)). It implies that constant strain-rate requirement is not necessary to meet the stress wave equilibrium condition.

Two plateau loadings were obtained using nylon shaper in a single SHPB test, as shown in Fig. 13(c) where stress wave equilibrium is achieved (see the light-green box). Again, stress wave non-equilbrium occurs when the incident wave increases rapidly (i.e. high loading rate) (see  time < 50 μs and 190 μs < time < 220 μs in Fig. 13 (c)).

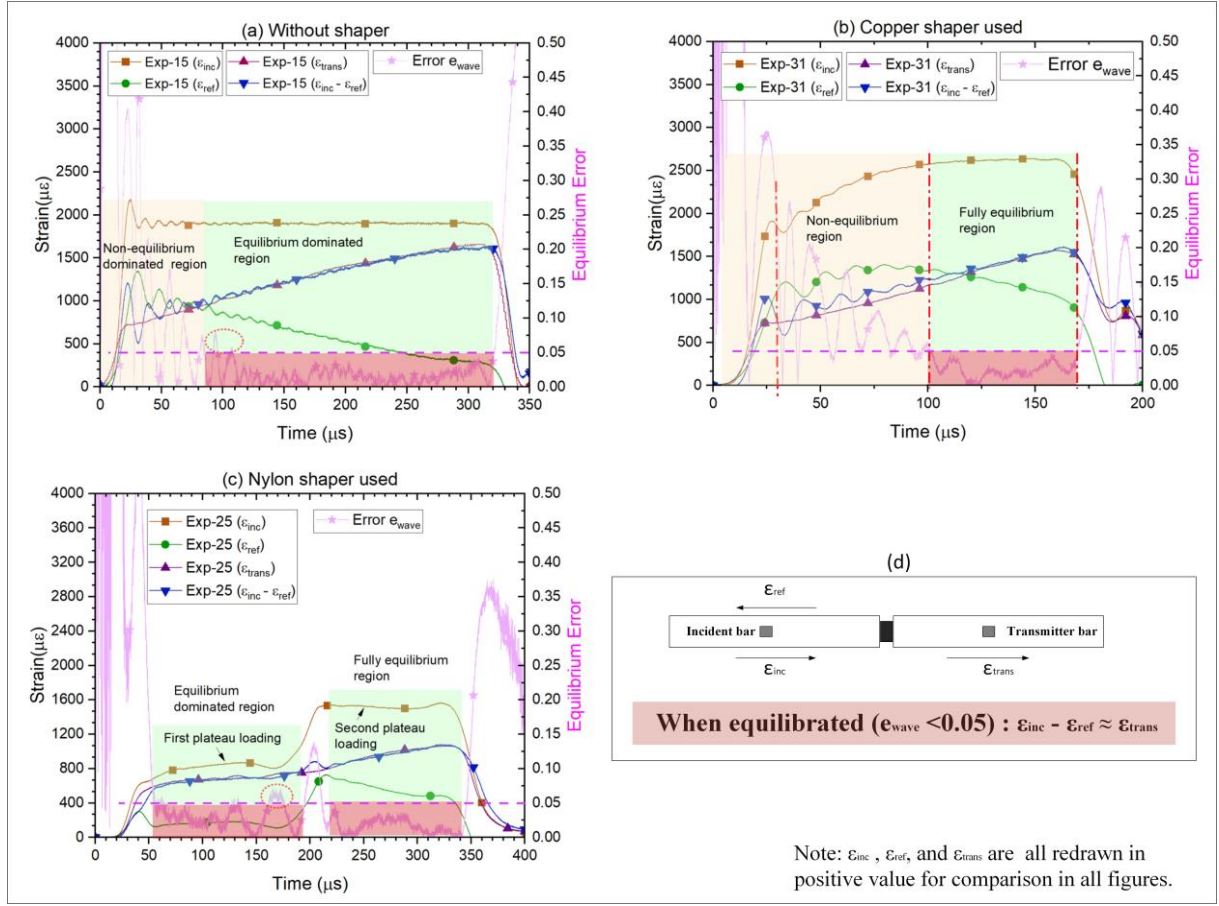

Fig. 13 Stress wave equilibrium analyses for SHPB tests: (a) Without shaper, (b) Copper alloy shaper, (c) Nylon shaper, (d) Diagram of the signals propagating in incident and transmitter bars.

### 4.2 Variation of stress with the instantaneous strain-rate in the specimen

Fig. 14 provides an example of the experimental results (Exp-18) for the variation of stress with strain and instantaneous strain-rate in a SHPB test when the nylon shaper with thickness of 2 mm is used. In Fig. 14(a), the stress increases with strain in overall. However, strain-rate show complicated variation with strain. Locally, unloading and reloading are observed at strain around 0.045 (see Fig. 14(c)) due to the use of shaper, causing reduced and even negative strain-rates. Fig. 14(d) presents corresponding time series data of the stress and strain-rate used in Fig. 14(a, c), which shows direct correlation between the unloading and reloading process with strain-rate.

To further evaluate the correlation between stress and strain-rate, over-stress is calculated. The over-stress is calculated by subtracting the stress at quasi-static strain-rate (0.0003 $s^{-1}$, Exp-Comp-1 in Fig. 7) from the dynamic flow stress given in Fig. 14(a). The results are provided in Fig. 14(b). It is evident that the variation of the instantaneous strain-rate causes the change of the instantaneous over-stress.

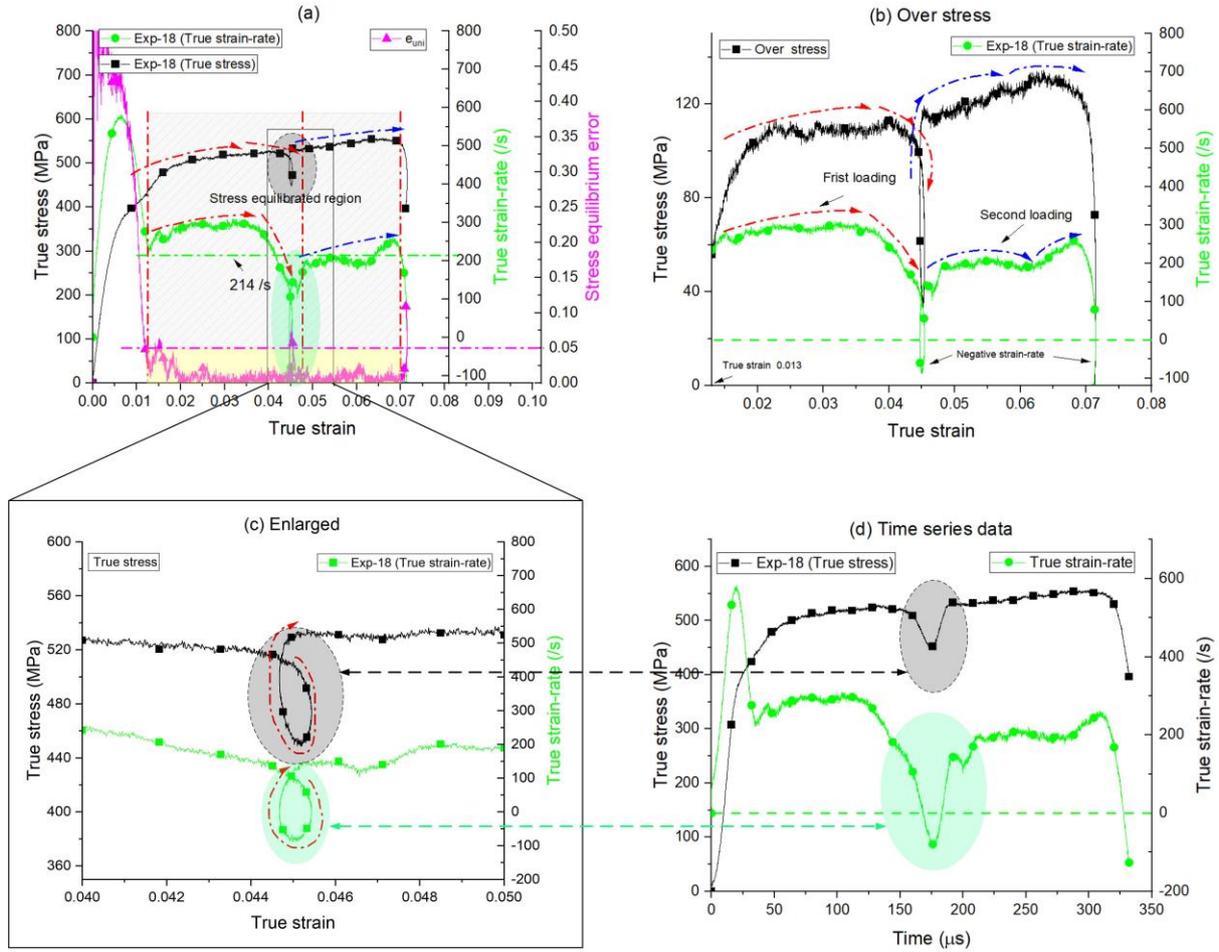

Fig. 14 Correlation between stress and strain-rate in a SHPB test with shaper (Exp-18): (a) Variations of stress and strain-rate with strain, (b) Variation of the over-stress (the difference between the dynamic flow stress and the quasi-static stress) and strain-rate with strain, (c) Enlargement of the local unloading and reloading data in (a), (d) Time series data of stress and strain-rate (Note: the true stress and true strain-rate are calculated by Eq.(5) and Eq.(2), respectively).

The stress equilibrium error, defined in Eq.(9), was calculated. If the acceptable stress equilibrium error is set as 5%, the stress equilibrium region can be obtained (see the light-grey and shaded box in Fig. 14(a)). In the stress equilibrium region, the averaged strain-rate is about 214 s$^{-1}$. If the flow stress is associated with the averaged strain-rate (214 s$^{-1}$), as it has been done conventionally, the dependence of the stress on the instantaneous strain-rate cannot be revealed, and consequently, the variation of the dynamic flow stress due to the variation of the instantaneous strain-rate cannot be explained. Therefore, the dynamic flow stress equation using averaged strain-rate in SHPB tests is not generally acceptable, especially when the strain-rate has relatively large variation in the stress equilibrium region (e.g. in Fig.14).

The variations of true stress with true strain and true strain-rate from two SHPB tests (i.e. Exp-27 without shaper and Exp-18 with shaper) are plotted in (stress, strain, strain-rate) space in Fig. 15(a, c) when different strain-rates (i.e. the instantaneous strain-rate and the averaged strain-rate) are used. The blue dots are the raw data of true stress obtained from the SHPB test by Eq.(5); the pink dots are the true stress associated with the averaged (constant) strain-rate according to the conventional method. The pink dots in Fig. 15(a, c) are the projection of the blue dots to the constant strain-rate plane.

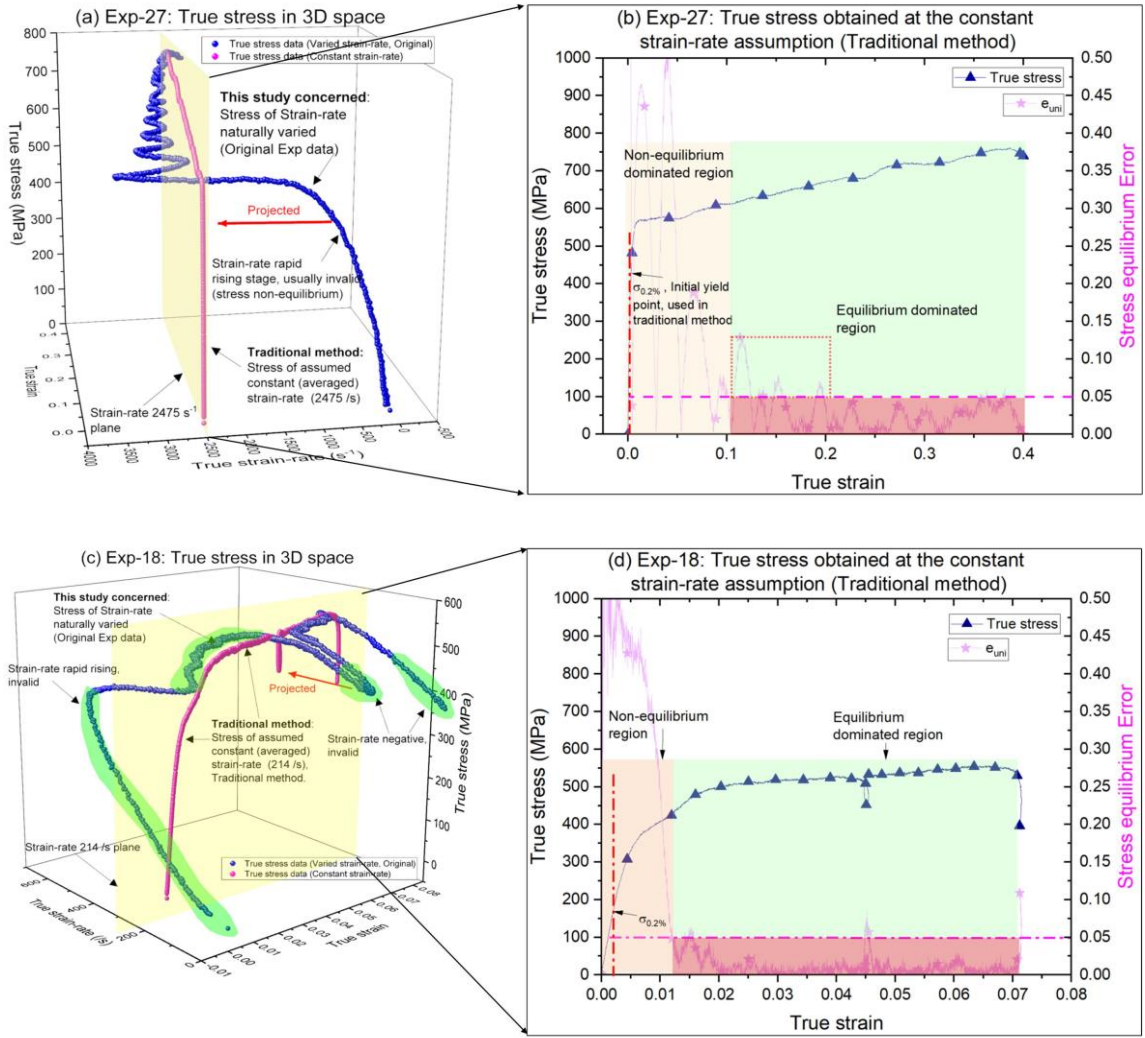

Fig. 15 Comparison of true stress-strain curves at instantaneous strain-rate and averaged strain-rate: (a, c) Exp-27 and Exp-18 data in (stress, strain, strain-rate) space, (b, d) True stress-strain curves of Exp-27 and Exp-18 at averaged strain-rate (Conventional method used).

The light-yellow regions in Fig. 15 (b, d) are the non-equilibrium regions, in which the true stress data are not qualified. Particularly, the true stress for true strain < 0.1=10% in Exp-27 and the true stress for true strain < 0.012=1.2% in Exp-18 are not qualified). However, such data (under the state of stress non-equilibrium) are frequently treated as valid data to determine the initial dynamic yield stress (e.g. $\sigma_{0.2\%}$ is often used to determine the initial dynamic yield stress whereas 0.2% strain is well within the region of stress non-equilibrium in both cases). Therefore, a great caution should be paid when SHPB test is used to determine the initial dynamic yield stress.

4.3  Data screening and ANN network training

In Section 3, a group of SHPB tests are conducted and preliminarily processed to obtain true stress, true strain and true strain-rate data shown in Fig. 12. As mentioned before, these data need to be checked before they are considered as qualified data to be used in the ANN training. Following two steps are applied to check and qualify the data.

(i) Step one: wave dispersion check

To investigate the effects of the wave dispersion and the alignment of pressure bars, special tests were designed. Nylon shaper (2mm thickness and 16 mm diameter) was used for Exp-16 and Exp-17 while no shaper was used for Exp-0.  In Exp-16, the transmitter bar was not installed in the test, and therefore, the incident wave was reflected totally without transmission. In Exp-0 and Exp-17, the incident and transmitter bars were directly connected without specimen. In this case, it was expected that the incident wave was transmitted to the transmitter bar totally without reflection.

The results are shown in Fig. 16. It is observed that the incident and reflected waves for Exp-16 are very close with error mostly smaller than 5% (see Fig. 16(a)), indicating negligible dispersion effect. For Exp-0 and Exp-17, the incident waves are transmitted into the transmitter bar with negligible wave reflections, as shown in Fig. 16 (b), indicating satisfactory contact and alignment between two pressure bars. Fig. 16 (b) also shows that the dispersion effect in transmitter bar is negligible. Thus, strain wave signals measured at strain gauge stations away from the specimen/bar interfaces can be directly used without dispersion corrections.

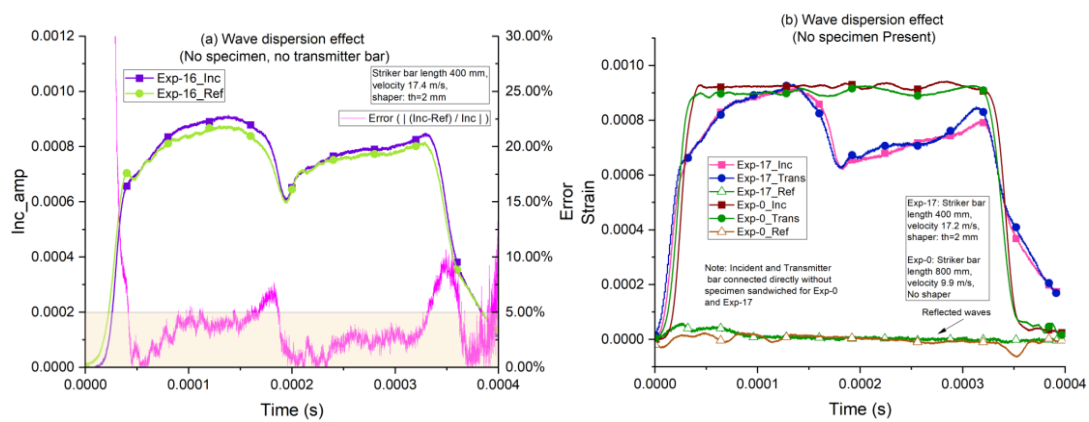

Fig. 16 Wave dispersion and distortion analyses, (a) No specimen and no transmitter bar (Exp-16); (b) Incident bar and transmitter bar connected directly without specimen (Exp-0 and Exp-17). Notes: the reflected waves are shifted and inversed for easy comparison,

(ii) Step two: data screening

Two typical sets of SHPB test data from Exp-15 (without shaper) and Exp-23 (with shaper) are used to demonstrate the data screening criteria presented in Section 2.1.2 (i.e. Eqs. (9) and (10)). The stress equilibrium error is set as 5 %, which has been widely accepted (e.g. [18, 19]).

As shown in Fig. 17(a), strong correlation between flow stress and strain-rate is observed, i.e. the instantaneous flow stress increases with the increase of instantaneous strain-rate, and vice versa. However, when the strain-rate abruptly changes, the flow stress on the rear surface of the SHPB specimen deviates instantaneously from that on the front surface because there is a time lag to transmit the suddenly changed stress in the SHPB specimen by stress wave propagation. It is evident that stress non-equilibrium occurs when strain-rate changes abruptly. The non-equilibrium data as well as negative strain-rate data in Fig. 17(a) are removed according to the criteria. The finally qualified data from Exp-15 is show in Fig. 17(b).

When shaper is used, the variation of strain-rate is more complicated. Three stress equilibrated regions (see yellow boxes in Fig. 17(c)) were obtained, where the equilibrium error is smaller than 5 %. The qualified stress data are presented in Fig. 17(d). It is also observed that, see the region between the plastic strain 0.1 and 0.2 in Fig. 17(c), the variation of strain-rate is nearly constant but the stress equilibrium condition does not meet. It means that constant strain-rate does not necessarily lead to the stress equilibrium between the two sides of the specimen.

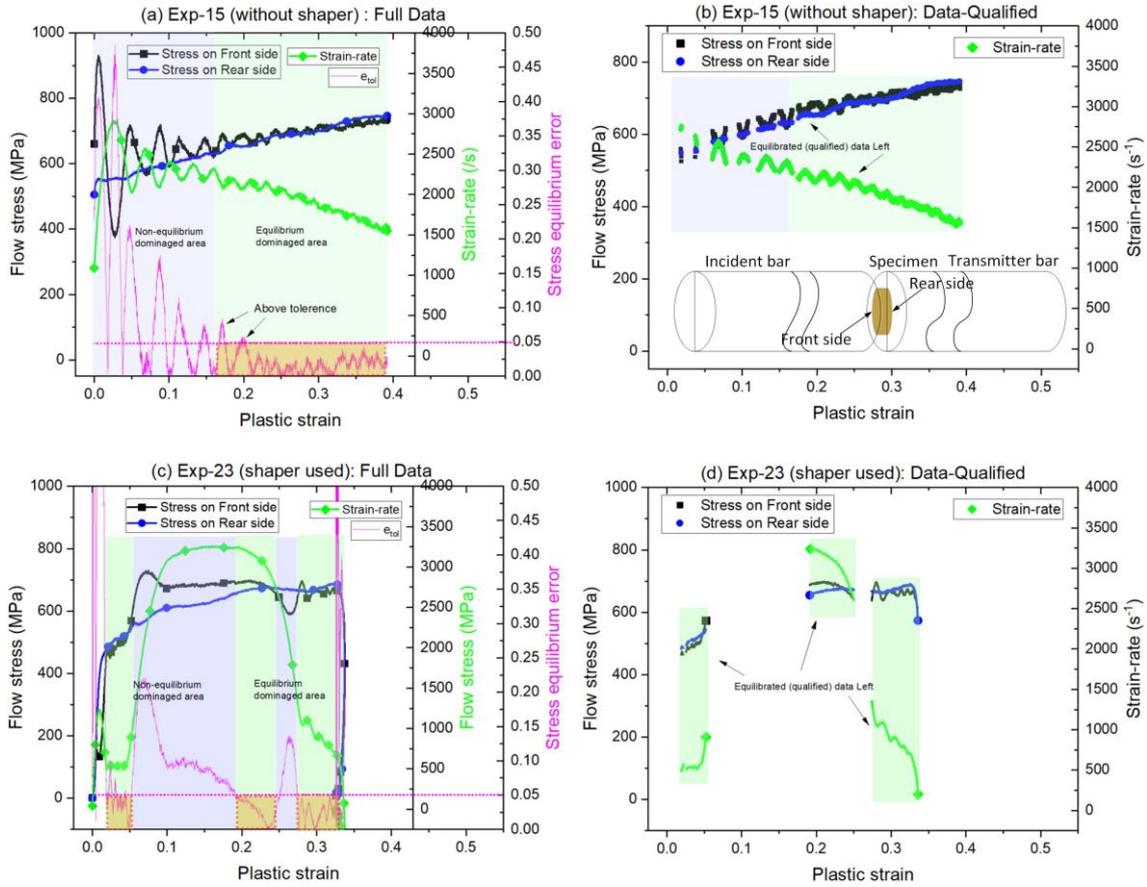

Fig. 17 (a, c) Full SHPB data of Exp-15 and Exp-27, (c, d) Qualified SHPB data of Exp-15 and Exp-27.

The same procedure is applied to analyse and screen all SHPB experimental data in Group A and Group B in Fig. 12. The qualified data for all SHPB tests are obtained and shown in Fig. 18, which are denoted as Data_Group_A (left column) and Data_Group_B (right column) respectively. In Fig. 18(c), Data_Group_A are well distributed in the measured strain and strain-rate space, roughly covering strain-rate range from 0 to 3700 s$^{-1}$ (quasi-static data included). The gently oscillated stress data shown in Fig. 18 are qualified, indicating that the stress at both ends of the SHPB specimen can be equilibrated when the strain-rate oscillates relatively slowly. The flow stress data in both Data_Group_A and Data_Group_B are qualified, but Data_Group_B are not usually adopted in conventional method in the determination of the dynamic flow stress.

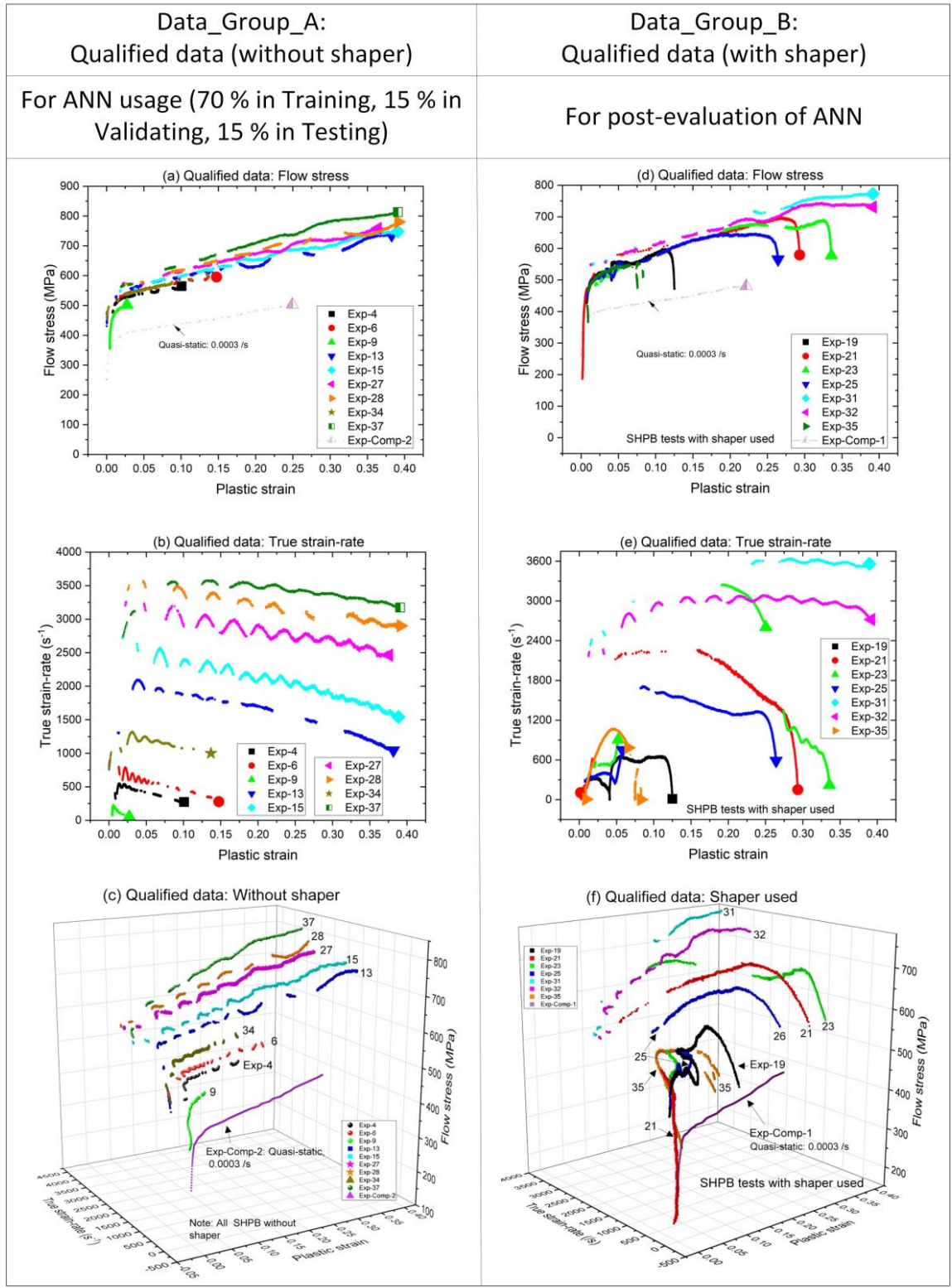

Fig. 18 Qualified data obtained from data screening: (a, b, c) Data_Group_A, (d, e, f) Data_Group_B.

Then, the qualified Data_Group_A are used to train ANN algorithm in the MATLAB/nnstart, which is a well-built machine learning kit. The input data are qualified plastic strain and strain-rate, and the target/output data are qualified flow stress. Divisions of data set for training, validating and testing are set as 0.70, 0.15 and 0.15, respectively, and the training function 'trainlm' in MATLAB/nnstart is used for every training network investigated in this study. When the network is trained by Data_Group_A, the trained ANN will be evaluated by Data_Group_B as the post-evaluation. Thus, Data_Group_B are independent SHPB tests that are not involved in ANN training in this study.

## 5 Validation and verification of the proposed method

### 5.1 ANN predicted results

Fig. 19 shows the MAPE of ANN (marked as ANN_MAPE) with different network structures trained by Data_Group_A. The ANN_MAPE is the error between the data (flow stress) predicted by trained ANN (marked as Network_Predicted_Data) and the experimental data in Data_Group_A (flow stress) under same strain and strain-rate coordinates. For simplicity, only one- and two-hidden layer networks are investigated. Similar results can be produced for other ANN types. Overall, the increase of neuron number per layer helps to improve the prediction accuracy. The MAPE tends to be stable as the neuron number in a layer is further increased. In addition, the overall performance for ANN with double hidden layer is much better than the performance of ANN with single hidden layer.

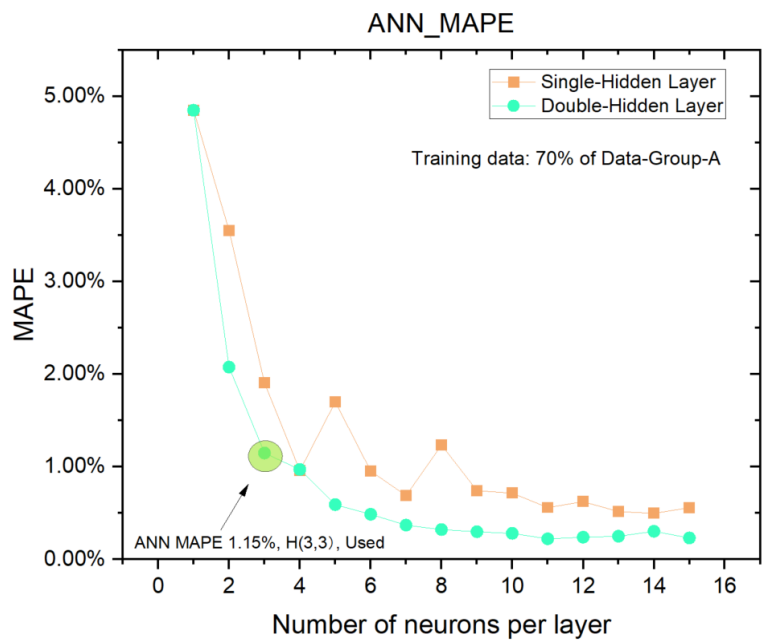

Fig. 19 The error with respect to the number of neurons per layer.

To decide the best network structure of double hidden layers, Network_Predicted_Data results of six networks are presented in Fig. 20 as a surface. The number of neurons in each layer ranges from one to six. The complexity (or non-linearity) of the Network_Predicted_Data surface increases with the increase of neuron number. The surface of H(1,1) network is quite flat, which approximately covers the most part of Data_Group_A, but cannot match the flow stress data at low strain-rate and low strain regions (see the red dashed region in Fig. 20(a)). When the number of neurons is increased to two (i.e. H(2,2)), only flow stress data at low strain region are unmatched (see the red dashed region in Fig. 20(b)). In Fig. 20(c), the Network_Predicted_Data by H(3,3) cover the Data_Group_A entirely without obvious deviations.

Although further increases of the neuron number to four or five can reduce ANN_MAPE, i.e. increasing the accuracy of Network_Predicted_Data, overfitting occurs (see the red dashed region in Fig. 20(d, e)). Overfitting becomes severer for H(6,6) network, which is unacceptable. Therefore, H(3,3) network is selected for the following analysis. The ANN_MAPE of H(3,3) network is 1.15 %.

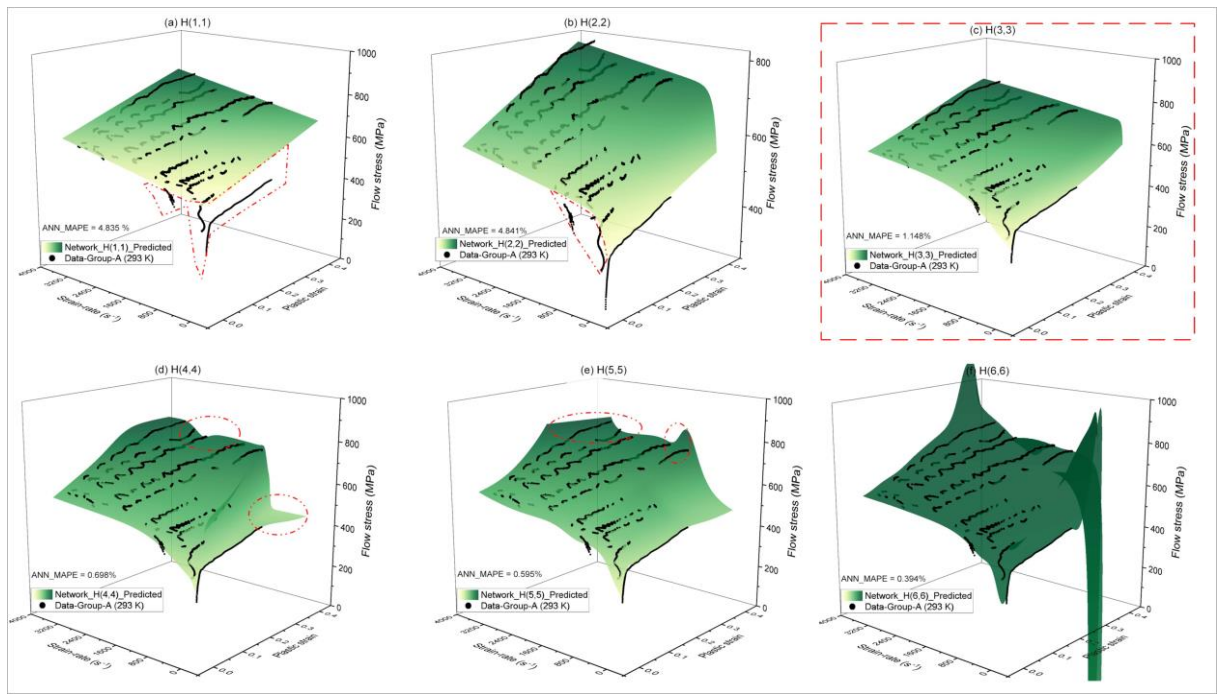

Fig. 20 Overfitting analysis: (a~f) j=1-6 number of neurons of double hidden layer network, e.g. network H(j, j) means that the network has two layers and the number of neurons in each layer is j.

The Network_H(3,3)_Predicted flow stress data and the experimental data involved in the training are shown in Fig. 21. Each SHPB test involved is indicated by discrete dots with a colour. Some abnormal data occur at initial deformation (see the red dashed circle) stage. Overall, remarkable agreement between experimental data and ANN-predicted data is observed in the whole strain and strain-rate space.

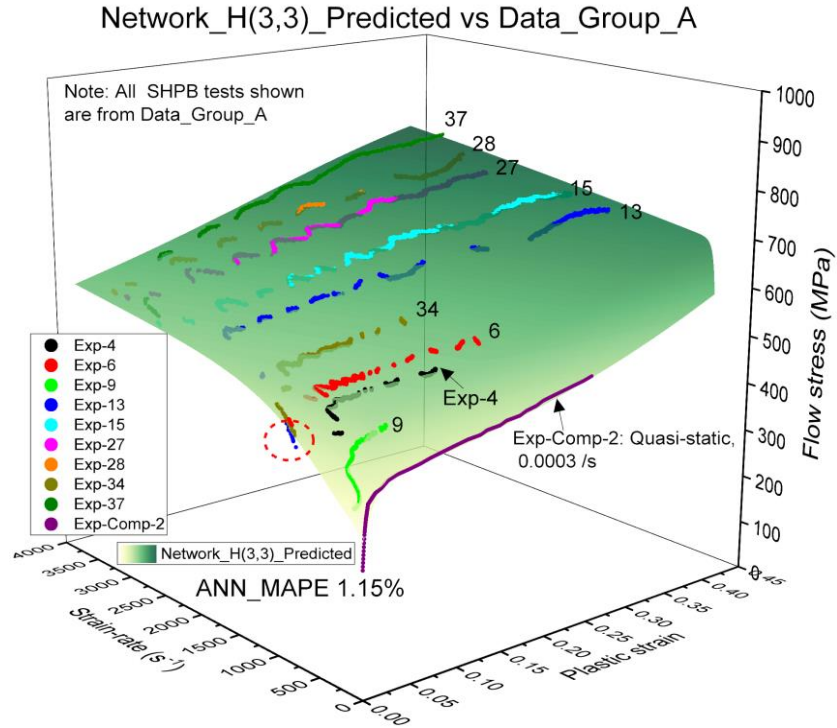

Fig. 21 Comparison of Network_H(3,3)_Predicted flow stress data and the experimental data involved in the training.

It is noted that, in this study, 70% of Data_group_A is randomly selected by the MATLAB/nnstart algorithm as the training data, while the rest 30% are for validating and testing. Hence, Data_group_A is the source data for the training, validating and testing of the ANN network algorithm used. The same procedure will be applied to the following newly constructed four sub-sets.

The effects of the source data (Data_Group_A) structure, i.e. the flow stress data distribution in strain and strain-rate variable space, on the performance of ANN predictions are investigated, which is crucial for the design of experiments in order to obtain the optimised data structure for the proposed method.

In the above training practice, the valid range of strain and strain-rate ($s^{-1}$) are in (0~0.4) and (0, 3700) as shown in Fig. 22. It is observed that the source data of training data in the (strain, strain-rate) space of (0, 0.4)×(0, 3700) ($s^{-1}$) are not evenly distributed, especially there are no data in the right lower corner when strain∈(0.15, 0.4) and strain-rate<1000 $s^{-1}$ (see Fig. 22(a))

To investigate the effect of the structure of source data on the performance of ANN, Data_Group_A are divided into four regions (see Fig. 22) simply based on the interval of strain-rates ($s^{-1}$): (0, 100), (100, 1000), (1000, 2500), (2500, 3700). Then, four sub-sets of Data_Group_A are constructed, i.e. Sub_Data_A1 (all Data_Group_A except those in Region-1), Sub_Data_A2 (all Data_Group_A except those in Region-2), Sub_Data_A3 (all Data_Group_A except those in Region-3), and Sub_Data_A4 (all Data_Group_A except those in Region-4).

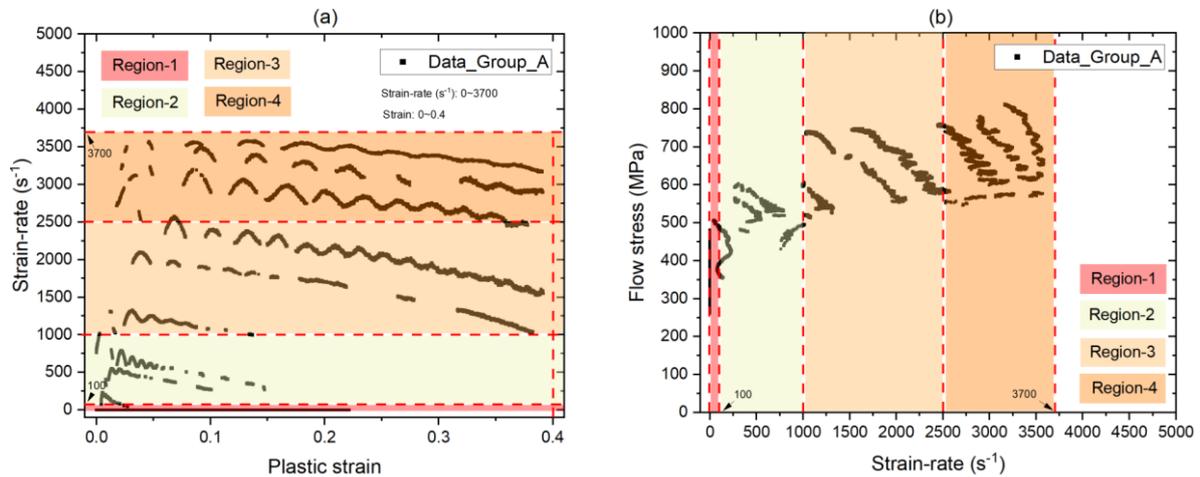

Fig. 22 Regions divided by strain-rate: (a) Strain-rate via strain, (b) Flow stress via strain-rate.

Then, ANN H(3,3) is trained individually using each of the above four sub-set source data. The results are shown in Fig. 23. It is shown that extrapolation ability of ANN outside the source data range is quite poor (see Regions A, D1 and D2 in Fig. 23(a, d)). Hence, source data boundary is important to improve the accuracy of ANN.

When the source data resolution is too low (i.e. the interval of any two consequent input points is too large), especially in the rapid transition regions, the results predicted by ANN are distorted compared with the experimental data (see Region B, C1 and C2 in Fig. 23(b, c)). It means that ANN may fail to predict the experimental data in the region with lower source data resolution even it is inside the source data boundary.

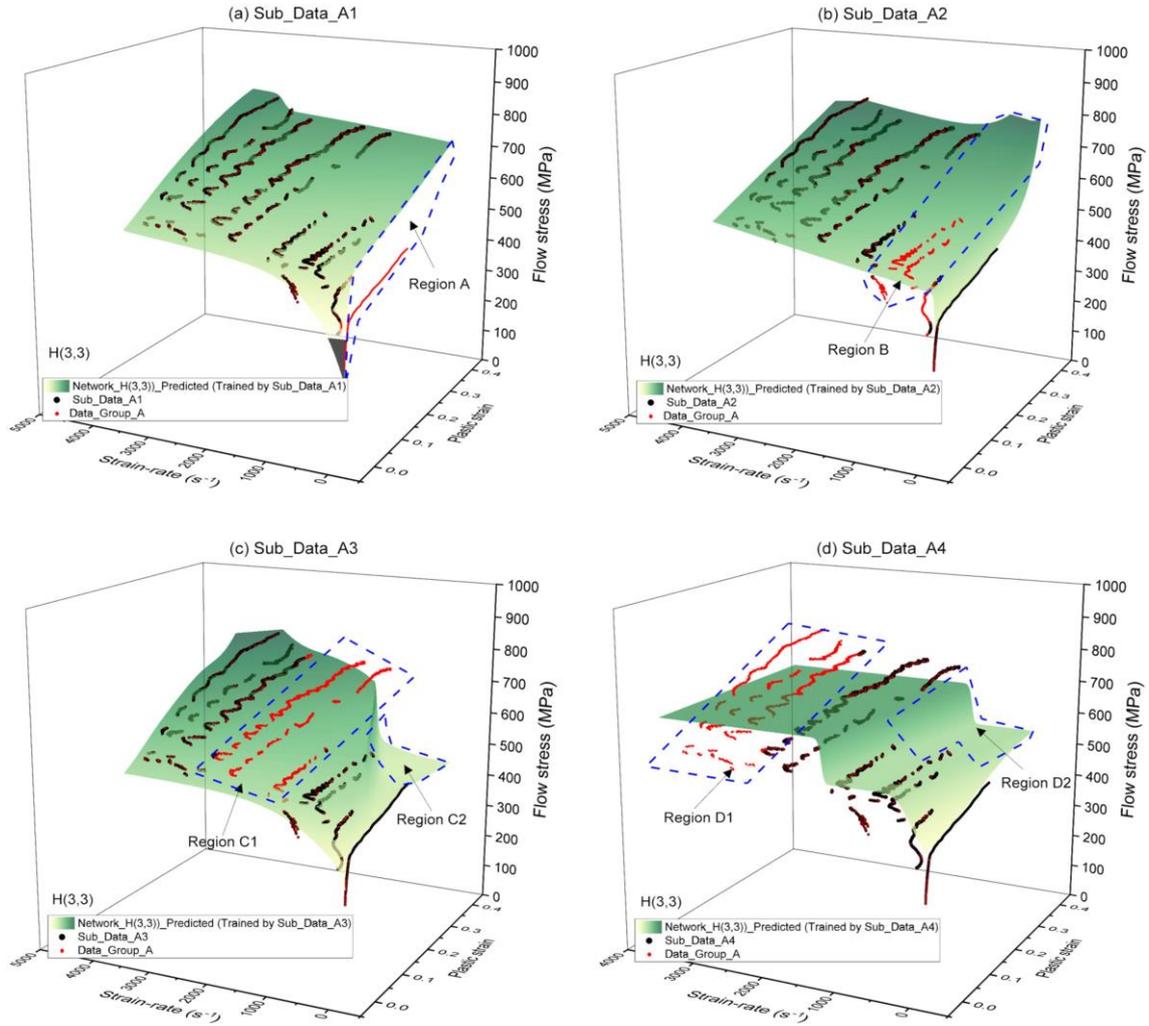

Fig. 23 The comparison of ANN-predicted data and all qualified data for (a) Sub_Data_A1, (b) Sub_Data_A2, (c) Sub_Data_A3, (d) Sub_Data_A4 (Note: All black dots belong to Data_Group_A, i.e. the black dots are covered by red dots).

To sum up, a well distributed experimental (source) data that meet following requirements are necessary for the better performance of the proposed method, i.e. (i) Data contain necessary features at transition points; (ii) The boundary of the source data of the ANN-predicted data; (iii) Source data resolution, measured usually by the largest interval of any two consecutive source data points, should be sufficiently high (i.e. the largest interval should be sufficiently small).

## 5.2 Post-evaluation of trained Network_H(3,3)

To further evaluate the trained Network_H(3,3) by Data_Group_A (denoted as Network_H(3,3)_Predicted), post-evaluation is conducted by a group of independent SHPB tests that not involved in any training. The qualified data of these SHPB tests are Data_Group_B as given in Fig.18 (d), (e) and (f) in Section 4.3. The comparison of Data_Group_B and Network_H(3,3)_Predicted data is shown in Fig. 24. Overall, Network_H(3,3)_Predicted data match well with the Data_Group_B.

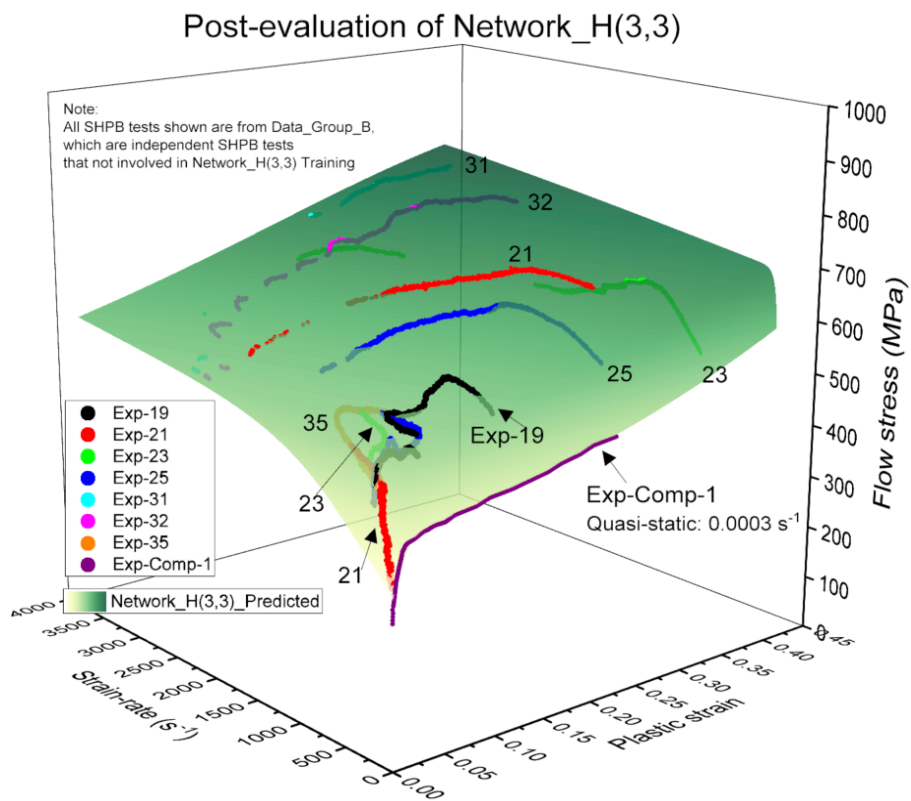

Fig. 24 Comparison of Data_Group_B and Network_H(3,3)_Predicted data. Note: Network_H(3,3)_Predicted was trained by the Data_Group_A. Here, Data_Group_B are independent SHPB tests not involved in Network_H(3,3)_Predicted training.

The quantitative comparison between Data_Group_B and Network_H(3,3)_Predicted data are presented in Fig. 25 and Fig. 26, in which errors of flow stress between these data are calculated in (strain, strain-rate) variable space, which are detailed below.

The qualified data of SHPB tests (Exp-23, -25, -31, and -32) from Data_Group_B are plotted in Fig. 25, together with their Network_H(3,3)_Predicted counterparts. The error between them is marked by violet triangular symbols. It is shown that the flow stress error is mostly smaller than 5%. There are some regions where the errors are larger than 5% (see Fig. 25(b, e, f)). Regions with relatively large errors are concentrated in the red dashed box (see

Fig. 25(c, d, e, f)). The Network_H(3,3)_Predicted data in these regions (see Fig. 25 (c, d)) are extrapolated data because no training and validation data exist in this region when Network_H(3,3) is trained. Therefore, the large error does exist in the region which has no training data available in a relatively large extent.

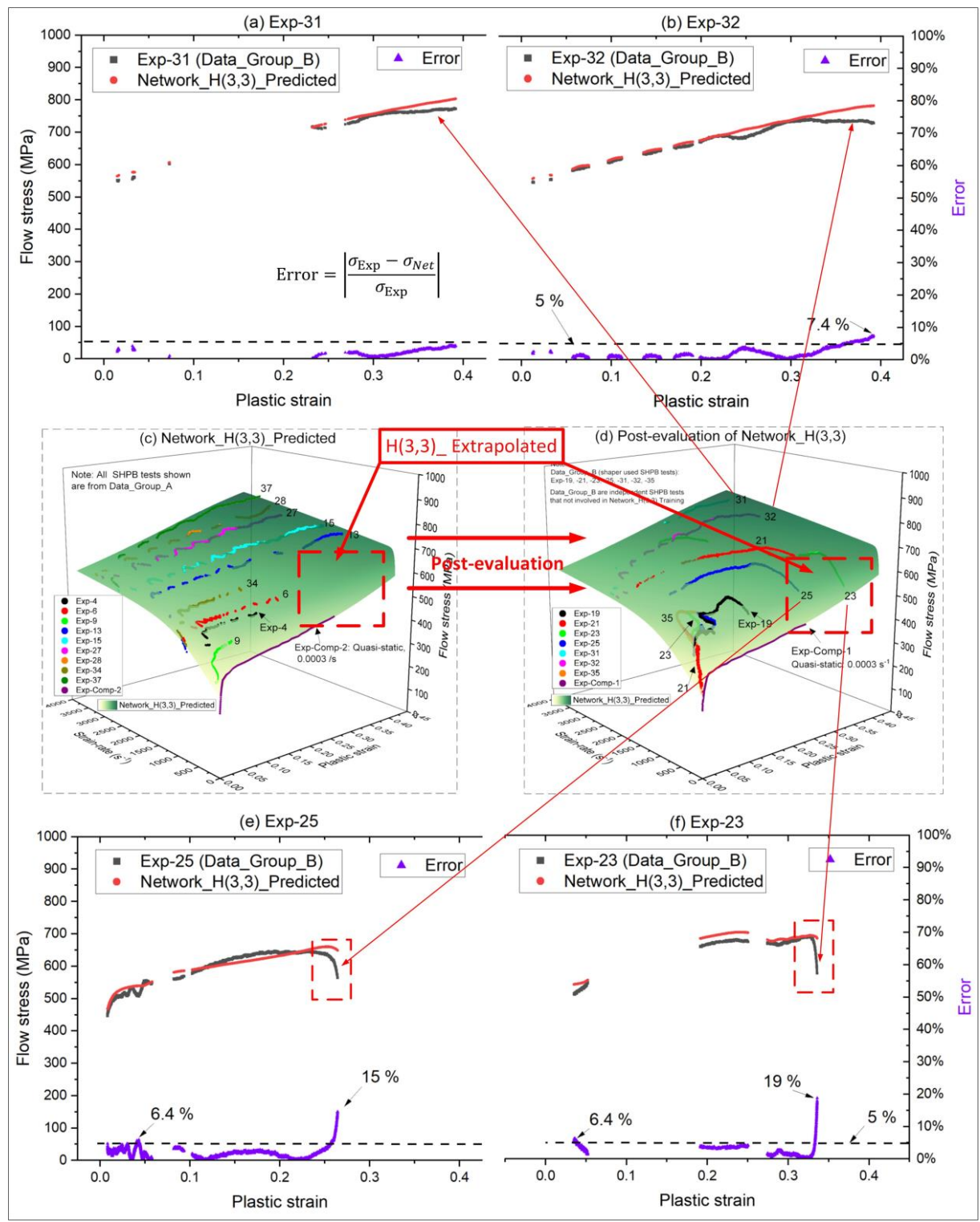

Fig. 25 Comparison of qualified data of SHPB tests (Exp-23, -25, -31, and -32) from Data_Group_B and Network_H(3,3)_Predicted data.

In Fig. 26, the qualified data of SHPB tests (Exp-19, -21, and -35) from Data_Group_B are outside the red dashed box marked in Fig. 25, but within the training data boundary. Because the training data (strain, strain-rate) variable space covers the (strain, strain-rate) variable space of (Exp-19, -21, and -35), the errors of the Network_H(3,3)_Predicted data for (Exp-19, -21, and -35) in Fig.26 are generally below 5%, which further proves the reliability and accuracy of trained Network_H(3,3).

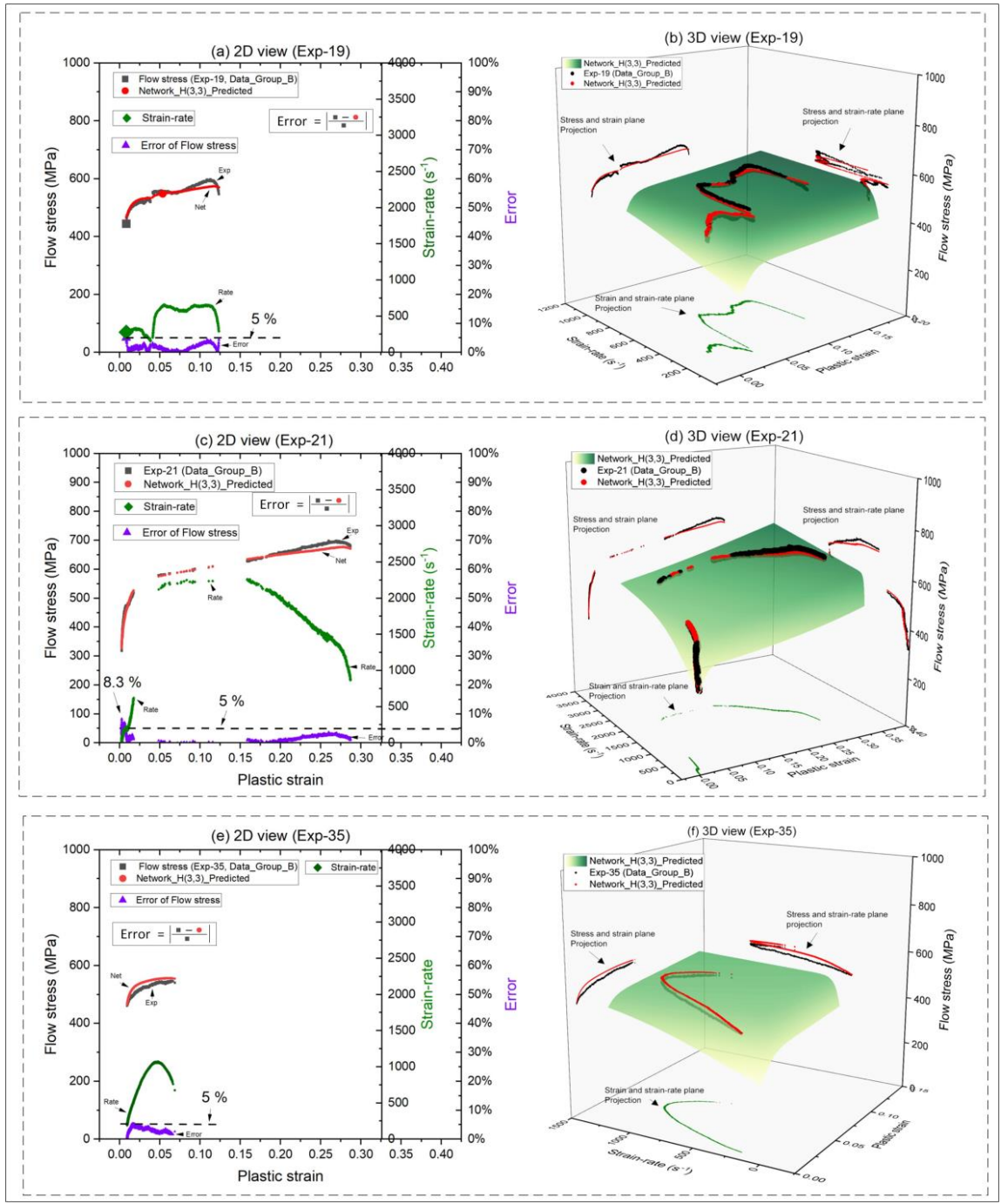

Fig. 26 Comparison of qualified data of SHPB tests (Exp-19, -21, and -35) from Data_Group_B and Network_H(3,3)_Predicted data.

From the above analyses, reasonably good agreement between Data_Group_B and Network_H(3,3)_Predicted data is achieved. Recall that the Network_H(3,3)_Predicted data deviates from Data_Group_A with ANN_MAPE=1.15% (see Fig. 21). Therefore, it is concluded that Data_Group_B is highly compatible with Data_Group_A.

It is noted that the Data_Group_B are from SHPB tests with shapers and the obtained true stress data in strain and strain-rate space are more complicated (see Fig. 12 (d, e, f)). The qualified data of these SHPB tests, i.e. Data_Group_B, are valid data although they are not used conventionally. If Data_Group_B is used as training data in Section 5.1 while Data_Group_A is used for post-evaluation of trained network, similar results can be obtained as shown in Fig. 27. The Network_H(3,3)_TB trained by Data_Group_B can well agree with the Data_Group_A (see Fig. 27(b)). It shows that there is no need to purposely achieve constant strain-rate (or use averaged strain-rate if constant/nearly-constant strain-rate cannot be achieved), which was generally practiced by previous researchers to determine the dynamic flow stress function from SHPB tests.

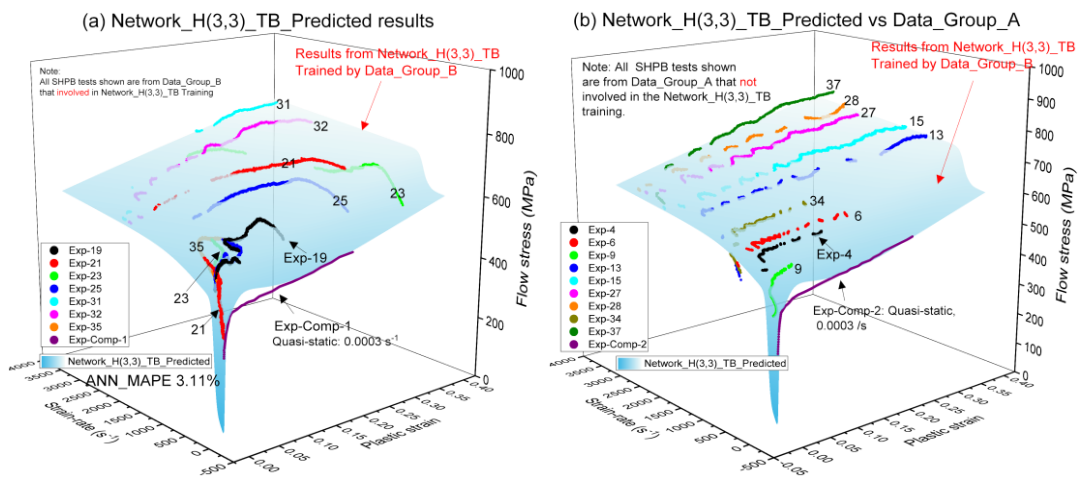

Fig. 27 (a) Network_H(3,3)_TB_Predicted results, (b) Post-evaluation of Network_H(3,3)_TB by Data_Group_A. Note: Network_H(3,3)_TB_Predicted is a ANN network trained by Data_Group_B. Here, Data_Group_A is independent data that not involved in the training of Network_H(3,3)_TB_Predicted.

5.3  Dynamic flow stress equation

Now, the trained Network_H(3,3) can predict whole-field flow stress in discrete form in the given space of strain and strain-rate, as shown in Fig. 21 (see Network_H(3,3)_Predicted). The trained Network_H(3,3) that generated these whole-field flow stress data can be implemented directly in the commercial numerical software (e.g. ABAQUS through its subroutine VUHARD [11]). However, such direct implementation of the trained Network_H(3,3) into a numerical code requires calculating the partial derivatives of the flow stress (the output of ANN) with respect to the strain and strain-rate variables when the subroutine VUHARD is called in every time step. The trained ANN contains dozens, if not hundreds, of weights and bias. The above issue can be resolved if an analytical flow stress equation or a low order fitting function based on the discrete flow stress generated by the Network_H(3,3) can be formulated. In this study, matrix decomposition method is used to obtain such function.

Firstly, the Network_H(3,3)_Predicted data is assembled as a FSD matrix **N,** as shown in Eq.(16). Using SVD, the matrix **N** can be decomposed to obtain decoupled flow stress relationship with strain and strain-rate variables (see Eq.(18)).

Fig. 28 shows the MAPE for different Rank-r approximation of SVD decomposition (see [7] for more detailed examples). It is seen that the MAPE of Rank-1 decomposition for the investigated material is 0.43 %, which is already quite small. MAPE tends to be zero for Rank-4 decomposition.

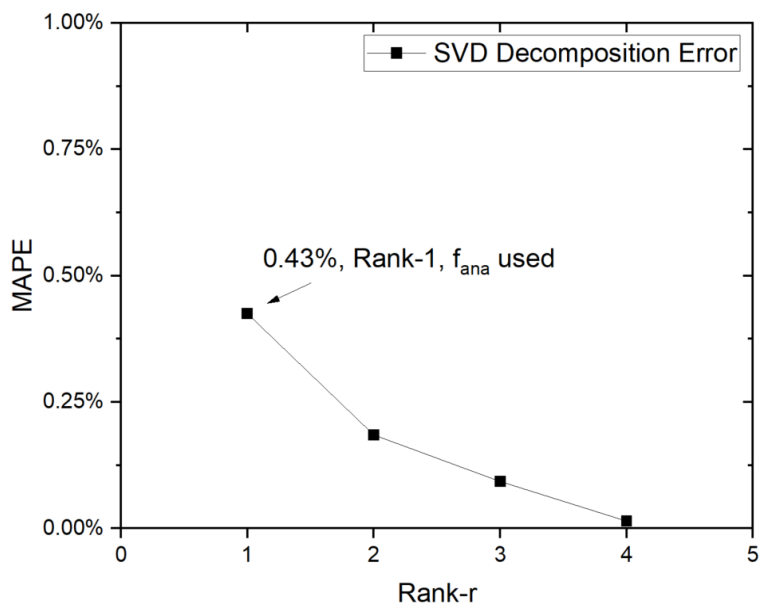

Fig. 28 MAPE for different Rank-r approximation.

The discrete flow stress equation from Rank-4 decomposition is

$$\sigma(\varepsilon, \dot{\varepsilon})_{Rank-4} = \sum_{k=1}^{4} f_{1,k}(\varepsilon) f_{2,k}(\dot{\varepsilon}) \quad (21)$$

The decomposition results of SVD Rank-4 are presented in Fig. 29. It is observed that $f_{1,1}(\varepsilon)$ owns the fundamental strain hardening behaviour of the investigated material, while $f_{1,k}(\varepsilon)$ tend to be zero when $k \geq 2$. For strain-rate behaviour, $f_{2,1}(\dot{\varepsilon})$ increases with strain-rate monotonically. The variation of $f_{2,k}(\dot{\varepsilon})$ when $k \geq 2$ is not monotonic.

The obtained flow stress equation given in Eq.(21) is in high precision. The implementation of this equation can simply use linear or spline interpolation method to fit the discrete data shown in Fig. 29.

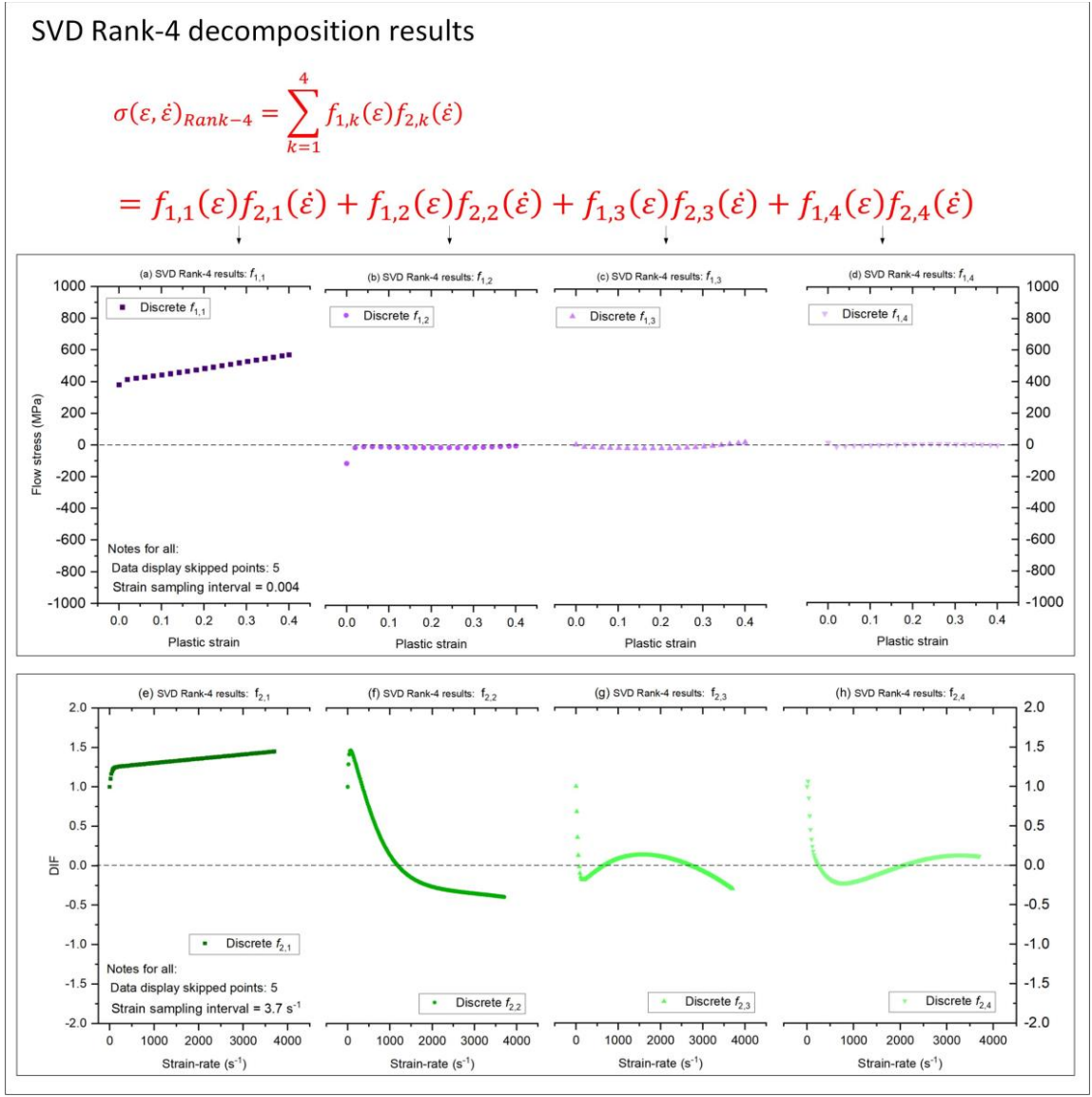

Fig. 29 Discrete data of SVD Rank-4 results: (a~d) $f_{1,k}(\varepsilon)$, (e~h) $f_{2,k}(\dot{\varepsilon})$, k =1, 2,3 4.

The accepted approximation error is set as 1.00% and Rank-1 approximation meets such error requirement, which means that the first term of SVD decomposition can be used to approximate **N** within this error. In this case, the error due to the use of Rank-1 decomposition is 0.43%.

Therefore, the approximated flow stress equation is given as follows

$$f(\varepsilon, \dot{\varepsilon})_{Rank-1} = f_{1,1}(\varepsilon) f_{2,1}(\dot{\varepsilon}) \tag{22}$$

where $f_{1,1}((\varepsilon)_i) = \lambda_1 \mathbf{u}_1(i)\mathbf{v}_1(1)$, $f_{2,1}((\dot{\varepsilon})_j) = \frac{\mathbf{v}_1(j)}{\mathbf{v}_1(1)}$; $\mathbf{u}_1(i)$ is the $i$th element of vector $\mathbf{u}_1$ and $\mathbf{v}_1(j)$ is the $j$th element of vector $\mathbf{v}_1$; $\mathbf{u}_1$, $\mathbf{v}_1$ and $\lambda_1 = 206790.25$ are SVD Rank-1 decomposition results. The discrete results of $f_{1,1}(\varepsilon)$ and $f_{2,1}(\dot{\varepsilon})$ by SVD Rank-1 decomposition are presented in Fig. 30.

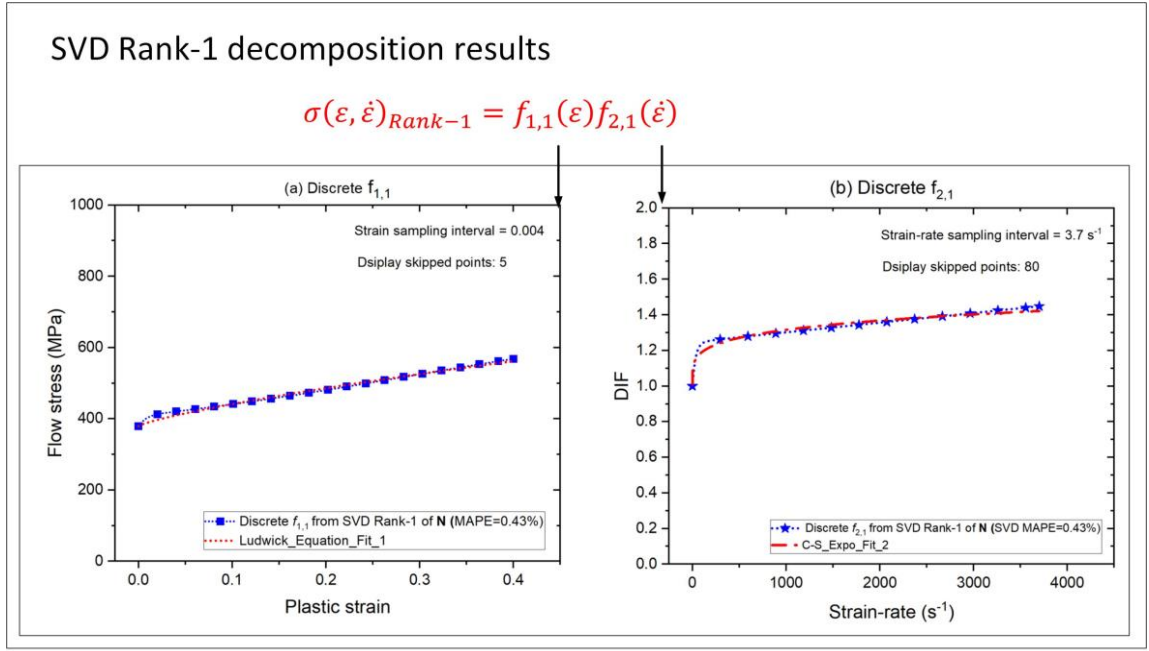

Fig. 30 SVD Rank-1 decomposition results: (a) $f_{1,1}(\varepsilon)$, and (b) $f_{2,1}(\dot{\varepsilon})$.

For better comparison, the flow stress from quasi-static compressive test (Exp-Comp-2) and DIFs at three plastic strains are plotted together, respectively, in Fig.31(a) and Fig.31(b). It shows that $f_{1,1}(\varepsilon)$ and $f_{2,1}(\dot{\varepsilon})$ are representatives of these experimental data as the 'eigen' strain hardening and 'eigen' dynamic increase factor (DIF) of the investigated material.

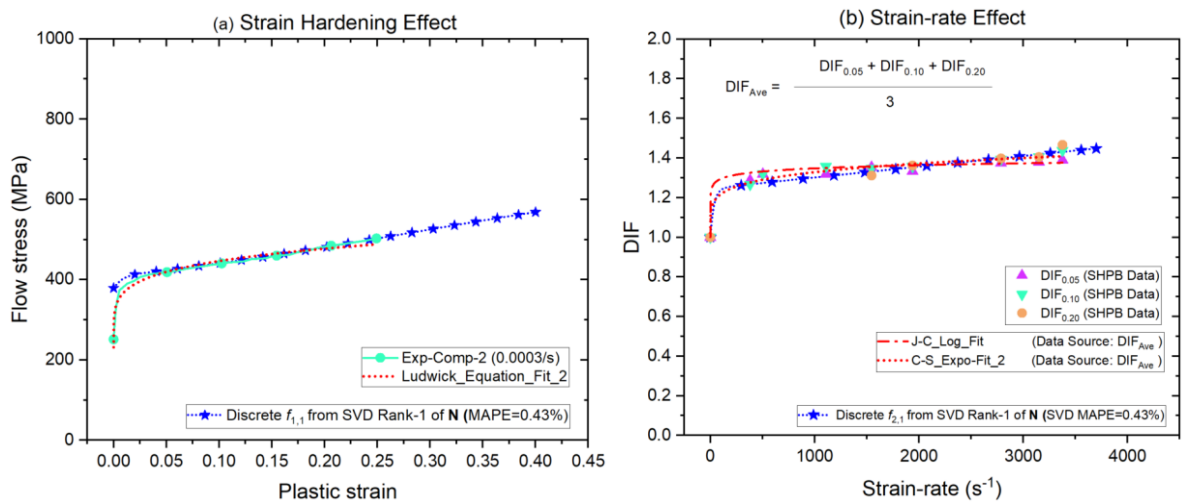

Fig. 31 Comparisons of conventionally-determined dynamic flow stress equation terms with Rank-1 decomposition results, (a) comparison with $f_{1,1}(\varepsilon)$ for strain hardening, (b) Comparison with $f_{2,1}(\dot{\varepsilon})$ for dynamic increase factor (DIF) (Note: The Data source of DIFs in Fig. 31(b) is given in Fig. 32(a))

The DIFs in Fig.31(b) are calculated according to the conventional method. The Data source of DIFs is presented in Fig. 32(a). An example of obtaining the averaged strain-rate for each SHPB test is presented in Fig. 32(b), which is required by the conventional method.

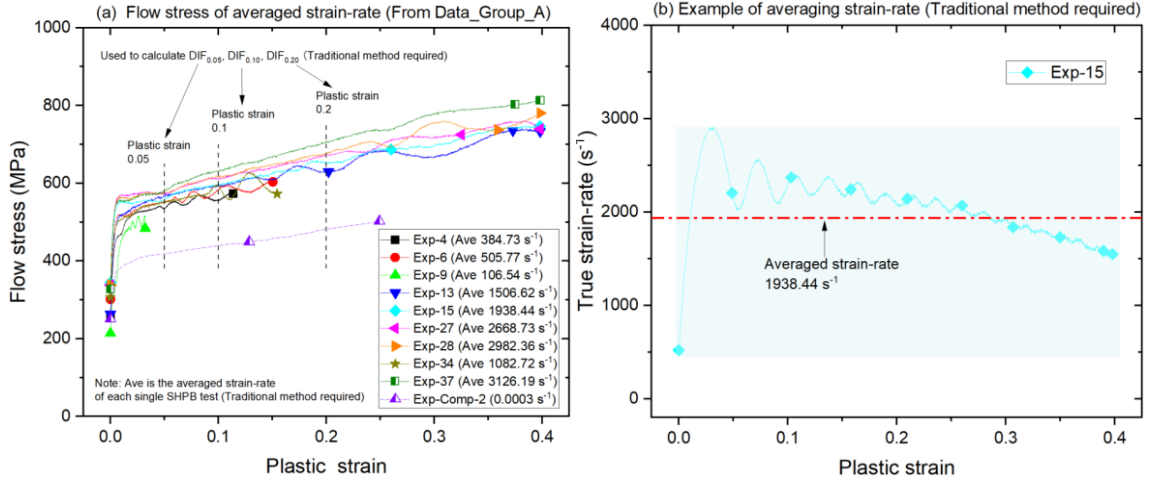

Fig. 32 (a) Data source used to obtain DIFs, and (b) an example of obtaining the averaged strain-rate for each test. Note: this figure is mainly presented for conventional method.

The discrete form of the flow stress equation given in Eq.(22) can be implemented directly through line/spline interpolation method in practical numerical coding stage.

As analytical form of flow stress equations is widely preferred in numerical modelling (but this is unnecessary as pointed out early), the data shown in Fig. 30 can be fitted with proper analytical functions through trial and error as an example, which may also help readers to better understand the proposed method.

Without loss of generality, an analytical dynamic flow stress equation for the investigated material is obtained as follows using Ludwik equation [20] (i.e. the first multiplicative function of J-C equation [6]) and Cowper-Symonds (C-S) equation [21], i.e.

$$f(\varepsilon, \dot{\varepsilon})_{ana} = f_{1,1}(\varepsilon) f_{2,1}(\dot{\varepsilon}) = (A_1 + B_1 \varepsilon^{n_1}) \left(1 + \left(\frac{\dot{\varepsilon}}{D_1}\right)^{\frac{1}{p_1}}\right) \quad (23)$$

where $A_1$, $B_1$, $D_1$, $p_1$, and $n_1$ are material constants obtained from fitting discrete $f_{1,1}(\varepsilon)$ and $f_{2,1}(\dot{\varepsilon})$ obtained in Fig. 31. The obtained constants of the analytical equation are given in Table A.3 in Appendix A. The fitting results are presented in Fig. 30 with good agreement.

The comparison of the $f(\varepsilon,\dot\varepsilon)_{ana}$ and experimental data is conducted in (stress, strain, strain-rate) space where Data_Group_A (involved in training, i.e. experimental dataset) and $f(\varepsilon,\dot\varepsilon)_{ana}$ surface (predicted datasets, coloured in green) are plotted in Fig. 33(a). The highly agreement of those two datasets prove the effectiveness of the proposed framework in the flow stress determination.

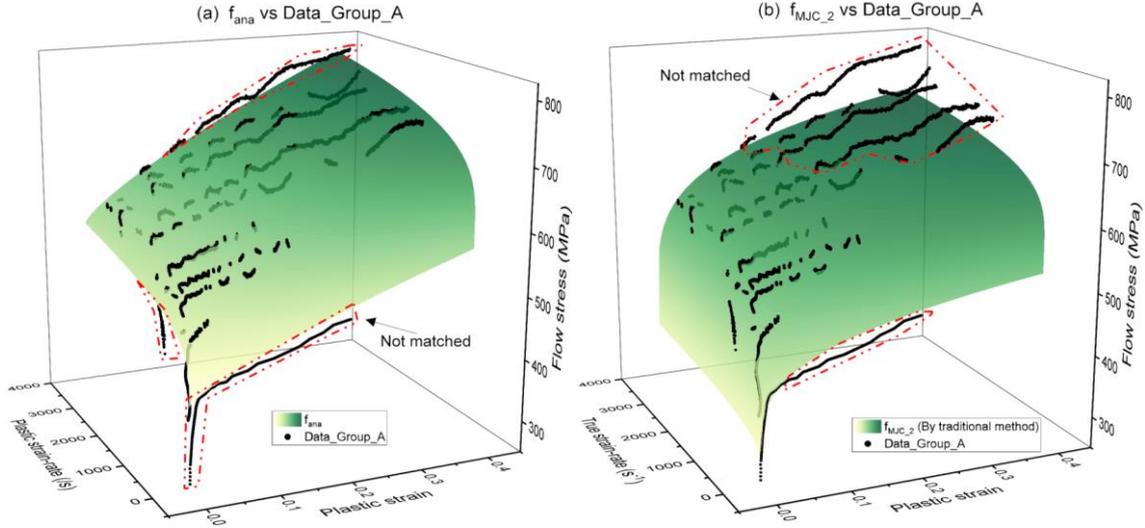

Fig. 33 Comparison of flow stress equations with experimental dataset, (a) $f(\varepsilon,\dot\varepsilon)_{ana}$ vs Data_Group_A, and (b) $f(\varepsilon,\dot\varepsilon)_{MJC}$ vs Data_Group_A.

It is also worth to simply compare the proposed method with conventional method (using empirical equation) in the determination of the flow stress equation. For conventional method, the flow stress at averaged strain-rate (Note: $1^{st}$ uncertainty) is used to obtain the DIF at a specified plastic strain. Then, the predefined flow stress equation (Note: $2^{nd}$ uncertainty) is set as $f(\varepsilon,\dot\varepsilon)_{MJC} = (A_2 + B_2\varepsilon^{n_2})\left(1 + \left(\frac{\dot\varepsilon}{D_2}\right)^{\frac{1}{p_2}}\right)$. The hardening effect at reference strain-rate is determined from Exp-Comp-2 (Note: $3^{rd}$ uncertainty) as shown in Fig. 31(a).

Three possible DIFs to characterize the strain-rate effect at three different plastic strains are presented in Fig. 31(b). It can be seen that the three DIFs are in different varying trend, thus averaged DIF (i.e. ($DIF_{0.05}$ + $DIF_{0.1}$ + $DIF_{0.2}$)/3) is used (Note: $4^{th}$ uncertainty). C-S equation [21] is introduced to fit the averaged DIFs ($5^{th}$ uncertainty). For comparison, J-C [6] logarithm term is also introduced to fit the averaged DIFs. The finally obtained material constants for $f(\varepsilon,\dot\varepsilon)_{MJC}$ is presented in Table A.4, in Appendix A. The direct comparison of $f(\varepsilon,\dot\varepsilon)_{MJC}$ and experimental data can be found in Fig. 33(b). It is observed that $f(\varepsilon,\dot\varepsilon)_{MJC}$ performs well when strain is smaller than 0.15. Beyond that, obvious difference between $f(\varepsilon, \dot{\varepsilon})_{MJC}$ and Data_Group_A is observed.

At least five uncertainties are generated, as noted above, for the use of the conventional method in the determination of dynamic flow stress equation. The flow stress equation ($f(\varepsilon, \dot{\varepsilon})_{MJC}$) obtained with J-C procedure is not reliable.

It is worth pointing out that the implementing of the proposed method is straightforward, and no multivariate functions are pre-requested. Although univariate functions are required in the fitting of SVD results ($\boldsymbol{f}_{1,1}$, and $\boldsymbol{f}_{2,1}$) (if an analytic flow stress equation is needed), it is simple (e.g. using univariate polynomial functions) and does not reduce accuracy.

## 6. Conclusions

A new methodology to determine dynamic flow stress of a metallic material is proposed in this study with focus on the effects of strain and strain-rate on the dynamic flow stress simultaneously. Based on systematic analyses of experimental data, following conclusions are obtained:

- Stress depends on strain-rate instantaneously. Using averaged strain-rate to replace instantaneous strain-rate in a SHPB test introduces error in the determination of dynamic flow stress.
- Data screening based on the stress equilibrium and non-unloading criteria is necessary and important to obtain the qualified source data for the development of either trained ANN or empirical dynamic flow stress equations.
- Constant strain-rate in a SHPB test is unnecessary; instead, varying strain-rate SHPB test is preferred to obtain rich data (as long as the data qualification criteria are satisfied).
- The data structure involved in the ANN training is important, i.e. a reduced and well-distributed data with sufficient resolution can save experimental cost, improve computational efficiency and maintain accuracy.

The results demonstrate the effectiveness of the proposed methodology in the determination of the dynamic flow stress in two-variable space. The general dynamic flow stress depends on three-variables when thermal effect is considered. The consideration of thermal effect on the dynamic flow stress will add an extra dimension in the variable space, which will increase the complexity of the methodology for the determination of the dynamic flow stress and the difficulty of the associated material tests. A detailed investigation on the determination of dynamic flow stress with considering strain, strain-rate and temperature variables will be presented in the Part 2 companion paper.

## Declaration of competing interest

The authors declare that no known competing interests exist in this study.

## Acknowledgements

The first author thanks Prof. Wei Zhang, Dr. Xiongwen Jiang and research students, Yu Tang, and Hongjian Wei for their support of the SHPB tests. The first author also thanks Dr. Yunfei Deng for the support of quasi-static compression tests. In addition, the first author acknowledges the financial support from The Henry Lester Trust.

# Appendix: Tables

Table A.1: The element composition of C54400.

| Element | Sn | Zn | Pb | P | Cd | Cu |
|---|---|---|---|---|---|---|
| Percentage (%) | 3.90 | 3.84 | 3.67 | 0.038 | 0.0003 | REM |

Table A.2: Experimental log of SHPB tests at room temperature (293 K).

| NO. | p (MPa) | $v_{st}$ (m/s) | $L_{st}$ (mm) | $D_{sh}$ (mm) | $L_{sh}$ (mm) | Material | Notes |
|---|---|---|---|---|---|---|---|
| Exp-0 | 0.6 | 9.9 | 800 | | | | *a |
| Exp-4 | 0.6 | 9.0 | 800 | | | | |
| Exp-6 | 0.6 | 10.0 | 800 | | | | |
| Exp-9 | 0.3 | 6.8 | 800 | | | | |
| Exp-13 | 1.5 | 16.6 | 800 | | | | |
| Exp-15 | 2.0 | 19.5 | 800 | | | | |
| Exp-16 | 0.6 | 17.4 | 400 | 16.0 | 2.0 | Nylon | *b |
| Exp-17 | 0.6 | 17.2 | 400 | 16.0 | 2.0 | Nylon | *a |
| Exp-18 | 0.6 | 17.4 | 400 | 16.0 | 2.0 | Nylon | |
| Exp-19 | 0.8 | 19.9 | 400 | 16.0 | 2.0 | Nylon | |
| Exp-21 | 1.2 | 25.7 | 400 | 10.0 | 3.0 | Nylon | |
| Exp-23 | 1.2 | 25.1 | 400 | 16.0 | 1.5 | Nylon | |
| Exp-25 | 1.2 | 25.3 | 400 | 16.0 | 3.0 | Nylon | |
| Exp-27 | 1.0 | 23.3 | 400 | | | | |
| Exp-28 | 1.2 | 25.3 | 400 | | | | |
| Exp-31 | 1.5 | 28.2 | 400 | 20.0 | 1.0 | Copper | |
| Exp-32 | 1.2 | 25.3 | 400 | 20.0 | 1.0 | Copper | |
| Exp-34 | 0.4 | 14.2 | 400 | | | | |
| Exp-35 | 0.4 | 14.1 | 400 | 20.0 | 1.6 | Paper | 16 pieces |
| Exp-37 | 1.5 | 28.0 | 400 | | | | |

Note: p is the gas pressure in the gas gun chamber used for driving the striker bar; velocity $v_{st}$ is the initial velocity of striker bar; $L_{st}$ is the striker bar length; $D_{sh}$ and $L_{sh}$ are the diameter and length of the shaper; *a: Incident and transmitter bar connected directly without specimen; *b: Transmitter bar is not installed, and specimen is not present.

Table A.3: Material constants of the analytic expression of the flow stress equation $f(\varepsilon, \dot{\varepsilon})_{ana}$.

|  | $A_1$ (MPa) | $B_1$ (MPa) | $n_1$ | $D_1$ | $p_1$ |
| --- | --- | --- | --- | --- | --- |
| $f_{1,1}(\varepsilon)$ | 379.20 | 374.11 | 0.78 | / | / |
| $f_{2,1}(\dot{\varepsilon})$ | / | / | / | 186038.19 | 4.53 |

Table A.4: Material constants of $f(\varepsilon, \dot{\varepsilon})_{MJC}$ by conventional method.

|  | $A_2$ (MPa) | $B_2$ (MPa) | $n_2$ | $D_2$ | $p_2$ |
| --- | --- | --- | --- | --- | --- |
| $f_{1,1}(\varepsilon)$ | 251.34 | 320.48 | 0.22 | / | / |
| $f_{2,1}(\dot{\varepsilon})$ | / | / | / | 495097.61 | 5.58 |